\documentclass[preprint,11pt]{elsarticle}
\usepackage[utf8]{inputenc}
\usepackage{caption}
\usepackage{subcaption}
\usepackage{graphicx}
\usepackage{amssymb}
\usepackage[tbtags]{amsmath}
\usepackage{mathtools}
\usepackage{hyperref}
\usepackage{geometry}
\usepackage{booktabs}
\usepackage{physics}
\usepackage{xcolor} 
\usepackage{float}
\usepackage{soul}

\definecolor{LightGray}{gray}{0.9}

\hypersetup{colorlinks=true, linkcolor=blue}

\date{\today}

\begin{document}

\begin{frontmatter}

\begin{abstract}
An individual's opinion concerning political bias in the media is shaped by exogenous factors (independent analysis of media outputs) and endogenous factors (social activity, e.g.\ peer pressure by political allies and opponents in a network). Previous numerical studies show, that persuadable agents in allies-only networks are disrupted from asymptotically learning the intrinsic bias of a media organization, when the network is populated by one or more obdurate agents (partisans), who are not persuadable themselves but exert peer pressure on other agents. Some persuadable agents asymptotically learn a false bias, while others vacillate indefinitely between a false bias and the true bias, a phenomenon called turbulent nonconvergence which also emerges in opponents-only and mixed networks without partisans. Here we derive an analytic instability condition, which demarcates turbulent nonconvergence from asymptotic learning as a function of key network properties, for an idealized model of media bias featuring a biased coin. The condition is verified with Monte Carlo simulations as a function of network size, sparsity, and partisan fraction. It is derived in a probabilistic framework, where an agent's opinion is uncertain and is described by a probability density function, which is multimodal in general, generalizing previous studies which assume that an agent's opinion is certain (i.e.\ described by one number). The results and their social implications are interpreted briefly in terms of the social science theory of structural balance.
\end{abstract}

\begin{keyword}
Media bias, Consensus, Partisans, Opinion dynamics, Opinion stability, Bayesian inference, Scale-free network
\end{keyword}

\title{Discerning media bias within a network of political allies: an analytic condition for disruption by partisans}

\journal{Physica A}

\author[add1]{Jarra Horstman\corref{mycorrespondingauthor}}
\cortext[mycorrespondingauthor]{Corresponding author}
\ead{jhorstman@student.unimelb.edu.au}

\author[add1,add2]{Andrew Melatos} 
\ead{amelatos@unimelb.edu.au}

\author[add3]{Farhad Farokhi} 
\ead{farhad.farokhi@unimelb.edu.au}

\date{\today}
\address[add1]{School of Physics, University of Melbourne, Parkville, VIC 3010, Australia}
\address[add2]{Australian Research Council Centre of Excellence for Gravitational Wave Discovery (OzGrav), University of Melbourne, Parkville, VIC 3010, Australia}
\address[add3]{Department of Electrical and Electronic Engineering, The University of Melbourne, Parkville 3010, Australia}

\end{frontmatter}

\newcommand{\opin}[3]{x_{#1}(#3,#2)}
\newcommand{\opinp}[3]{p'_{#1}(#2,#3)}
\newcommand{\opinc}[2]{\opin{\mathrm{cons}}{#2}{#1}}
\newcommand{\nodes}{\mathcal{N}}
\newcommand{\pnodes}{\mathcal{N}_{\mathrm{p}}}
\newcommand{\rnodes}{\mathcal{N}_{\mathrm{r}}}
\newcommand{\edges}{\mathcal{E}}
\newcommand{\bias}{\theta_0}
\newcommand{\ptheta}{\theta_{\mathrm{p}}}
\newcommand{\pthetai}[1]{\theta_{\mathrm{p},#1}}
\newcommand{\lht}[2]{P[S(#1)|#2]}
\newcommand{\topic}{\Theta}
\newcommand{\sumset}[2]{\sum_{#1 \in #2}}
\newcommand{\bel}[2]{\pi_{#1}(#2)}
\newcommand{\pbel}{\pi_{\mathrm{p}}}
\newcommand{\belp}[2]{\pi'_{#1}(#2)}
\newcommand{\bfac}[1]{\Lambda(#1)}
\newcommand{\bfunc}[2]{f[#1,#2]}
\newcommand{\stat}{\hat{\pi}}
\newcommand{\inflim}[1]{\lim_{#1\rightarrow\infty}}
\newcommand{\bern}{B(\bias)}
\newcommand{\kldiv}[2]{\mathrm{KL}\big\{#1||#2\big\}}
\newcommand{\kldivbern}[1]{\kldiv{\bern}{\lht{t}{#1}}}
\newcommand{\plam}{\lambda_{\mathrm{p}}}
\newcommand{\plapl}{\vb{L}'_{\mathrm{p}}}
\newcommand{\belc}[1]{\pi_{\mathrm{cons}}(#1)}
\newcommand{\belcs}{\hat{\pi}_{\mathrm{cons}}}
\newcommand{\pdeg}[1]{d_{#1,\mathrm{p}}}
\newcommand{\rdeg}[1]{d_{#1,\mathrm{r}}}

\section{Introduction}
\label{sec:intro}
The public perception of media organizations as politically biased is a fact of life \cite{baron_persistent_2006,groeling_media_2013}. Media bias can skew public opinion, alter electoral outcomes \cite{druckman_impact_2005,eberl_one_2017}, or affect public health \cite{han_how_2022,viswanath_mass_2007}. The influence of the media is felt directly, e.g.\ when individuals consume biased media outputs \cite{druckman_impact_2005,eberl_one_2017}. It is also felt indirectly, when individuals self-regulate how receptive they are to messages from a particular organization after assessing the degree to which the bias accords with their own, native political orientation \cite{eberl_one_2017,eveland_impact_2003,perloff_three-decade_2015}. Assessing the bias of a media organization is intrinsically a social activity. It occurs exogenously, when individuals analyze media outputs independently, e.g.\ by reading regularly the editorials of a newspaper \cite{druckman_impact_2005,eberl_one_2017}. It also occurs endogenously, when individuals share their beliefs about the bias itself (as distinct from the underlying economic and political issues) across a network of political allies and opponents, e.g.\ via social media \cite{eveland_impact_2003,southwell_roles_2007,ioannides_job_2004}. The interplay between independent observation and peer pressure leads to several complex and counterintuitive phenomena, such as converging more quickly to false perceptions than true ones or never converging to a stable perception at all \cite{fang_opinion_2020,low_discerning_2022,low_vacillating_2022,bu_discerning_2023}. \\

Opinion dynamics models are a versatile tool for investigating how a network of politically affiliated agents collectively form perceptions about media bias \cite{low_discerning_2022,low_vacillating_2022,bu_discerning_2023,carletti_how_2006,pineda_mass_2015,taylor_towards_1968,martins_mass_2010}. Such models can be classified according to how an agent's opinion is represented, what information causes the opinion to evolve, and how the evolution occurs, e.g. through either Bayesian or non-Bayesian learning. Starting with the work of French and DeGroot \cite{french_formal_1956,degroot_reaching_1974}, deterministic, scalar models have been formulated, in which an agent's opinion is described by a single real number; that is, the agent holds a single belief with certainty. An agent's opinion is modified by peer pressure in a non-Bayesian fashion via a weighted average of the opinions of neighbors in the network; that is, the update rule is linear \cite{proskurnikov_tutorial_2017,proskurnikov_tutorial_2018}. A media organization can be inserted into this scalar, linear framework as a special additional agent, whose opinion is static and contributes to peer pressure in the network without being affected itself \cite{carletti_how_2006,pineda_mass_2015,taylor_towards_1968,martins_mass_2010}. Such studies find that the media is most influential when it is accessible, adheres closely to other opinions in the network, and broadcasts to persuadable agents rather than obdurate agents (partisans) \cite{carletti_how_2006,pineda_mass_2015}. \\

In many applications, agents are uncertain about their beliefs; that is, they entertain multiple opinions simultaneously, to which they assign probabilities. This is important especially when partisans disrupt the opinion formation process. Multi-dimensional opinions are usually represented by a vector \cite{tian_dynamics_2023,parsegov_novel_2017,demarzo_persuasion_2003} or a probability distribution function (PDF) \cite{low_discerning_2022,anunrojwong_naive_2018,jadbabaie_non-bayesian_2012}. They are an essential part of modeling perceptions about media bias. For example, a newspaper's editorials may occupy different locations on a left-right spectrum when addressing social versus economic issues, creating uncertainty among readers as to whether the newspaper is ``truly" left- or right-wing. The use of a PDF lets agents be modeled as Bayesian learners, who adjust their prior opinions in response to data interpreted through a likelihood function (i.e.\ their internal model of a broadcast media signal) \cite{low_discerning_2022,anunrojwong_naive_2018,jadbabaie_non-bayesian_2012,fang_social_2019,fang_opinion_2020,lalitha_social_2016,mossel_efficient_2016}. Bayesian learning usually occurs in response to independent observations of an exogenous signal, e.g.\ when an agent reads a newspaper editorial themselves. In contrast, the endogenous second step, in which an agent shares their posterior opinion following an independent observation with political allies and opponents in the network, occurs in either a non-Bayesian \cite{low_discerning_2022,low_vacillating_2022,bu_discerning_2023,jadbabaie_non-bayesian_2012,fang_social_2019,lalitha_social_2016} or Bayesian manner \cite{anunrojwong_naive_2018,fang_opinion_2020,mossel_efficient_2016}\footnote{The existence of different learning mechanisms for endogenous information transfer is justified in Ref. \cite{demarzo_persuasion_2003,jadbabaie_non-bayesian_2012,rahimian_bayesian_2016}. Weighted averaging \cite{degroot_reaching_1974} can be interpreted as a form of naive Bayesian learning \cite{jackson_naive_2007,demarzo_persuasion_2003} or as a utility optimization problem \cite{rahimian_bayesian_2016}. This latter interpretation also leads to weighted geometric averaging \cite{rahimian_bayesian_2016,lalitha_social_2016}.}. A rich and counterintuitive variety of long-term behaviors emerge from such probabilistic models. Sometimes agents learn the ground truth about media bias irrespective of internal models and network topology \cite{jadbabaie_non-bayesian_2012}. Sometimes agents fail to learn the ground truth by never converging on an answer \cite{low_discerning_2022,low_vacillating_2022,bu_discerning_2023,fang_social_2019,lalitha_social_2016} or converging to a wrong answer \cite{low_discerning_2022,low_vacillating_2022,bu_discerning_2023,anunrojwong_naive_2018}. The behavior arises from a complex interplay between the explanatory power of an agent's internal model and how the network topology disseminates information. \\

The behavior predicted by probabilistic models of media bias is complex with or without partisans \cite{low_discerning_2022,low_vacillating_2022,bu_discerning_2023}. Characterizing it requires extensive Monte Carlo simulations, which are expensive computationally; those in Ref. \cite{low_discerning_2022,low_vacillating_2022,bu_discerning_2023} are restricted to $\lesssim 10^2$ agents, for example. Moreover, it is challenging to identify general conditions that give rise to the observed behavior, because there are many moving parts in each simulation: priors, signal sequence, network topology, and so on. In this paper, we investigate how to approximate the model in Ref. \cite{low_discerning_2022,low_vacillating_2022,bu_discerning_2023} in terms of a simplified two-state model, in which each agent holds an opinion about just two possible values of the bias instead of a continuous spectrum of values (cf.\ 21 states in Ref. \cite{low_discerning_2022,low_vacillating_2022,bu_discerning_2023}). We derive analytically a general condition for when the agents succeed in asymptotically learning the intrinsic bias of a media organization despite the disruptive influence of partisans and when they do not --- the central result of the paper. The paper is organized as follows. In Section \ref{sec:model}, we discuss the idealization of the media bias inference problem in terms of a biased coin and review the probabilistic opinion update rule introduced in Ref. \cite{low_discerning_2022} and its generalization to include obdurate partisans \cite{bu_discerning_2023}. \\

In Section \ref{sec:two-state}, we introduce the two-state approximation, reformulate the update rule as a system of nonlinear difference equations, and solve for the stationary points. In Section \ref{sec:stability}, notions of stability are introduced, and a general analytic condition is presented, for when persuadable agents are disrupted by partisans from asymptotically learning the coin bias. The properties of the instability condition are investigated analytically and two modes of partisan disruption are distinguished. In Section \ref{sec:montecarlo}, the instability condition is verified numerically with Monte-Carlo simulations as a function of the size and sparsity of the network, and the fraction of partisans in the network. The results and their social impact are interpreted briefly in terms of the social science theory of structural balance in Section \ref{sec:conclusion}.

\section{Inferring the bias of a coin in a network: an idealized model of media bias}
\label{sec:model}
The behavior of a network of political allies and opponents, as they strive to infer the political bias of a media organization, is complicated by many hard-to-quantify human factors, including the innate psychology of individuals \cite{eveland_impact_2003}, the social norms and cognitive frameworks underpinning information processing by groups \cite{eveland_impact_2003,perloff_three-decade_2015,lee_liberal_2005,southwell_roles_2007}, and the inadvertent corruption or loss of information through imperfect communication \cite{southwell_roles_2007,carlson_through_2019}. In addition, there are indications from human experiments, that agents in a network employ two distinct modes of reasoning: one that is non-deliberative, habitual, Markovian and occurs on fast time-scales, and another that is deliberative, conscious, strategic, non-Markovian and occurs on slow time-scales \cite{evans_two_2003,kahneman_thinking_2011}. In this paper, we disregard the above complications and many others and analyze instead an idealized version of the real-life media bias problem, in which a network of agents strive to infer the true bias of a coin by making independent observations exogenously and sharing opinions endogenously, the latter mechanism being a type of peer pressure. The coin's bias, expressed through a sequence of coin tosses, is analogous to a newspaper's political stance on a left-right spectrum, expressed through a sequence of editorials, for example. The agents maintain pairwise political relationships (allies or opponents), which control in part how opinions about the media organization's bias diffuse through the network by peer pressure, when agents share their posterior beliefs at every time step. In Section \ref{subsec:coin}, we define the public signal broadcast by a biased coin. In Section \ref{subsec:step1}, we codify how agents independently update their opinion about the bias by observing the coin toss and applying Bayes's rule. In Section \ref{subsec:step2}, we define how peer pressure in the network is modeled with a non-Bayesian, linear averaging rule. To keep the exposition clear, we present the theory in Sections \ref{subsec:coin}--\ref{subsec:step2} in terms of an arbitrary network containing allies and opponents. However, the analytic calculations in Sections \ref{sec:two-state} and \ref{sec:stability} apply specifically to allies-only networks; we do not know how to generalize them to opponents-only and mixed networks at the time of writing, for reasons explained in Sections \ref{subsec:updrule} and \ref{sec:conclusion}. In Section \ref{subsec:conv}, we discuss how persuadable agents in allies-only networks achieve consensus and learn asymptotically \cite{low_discerning_2022}. In Section \ref{subsec:partisans}, we review an extension of the update rule to include obdurate partisans, and review their disruptive effect on an allies-only network, studied previously \cite{bu_discerning_2023}.
\subsection{Sequence of coin tosses}
\label{subsec:coin}
Let us assume that a media organization publishes regularly a politically relevant output (e.g.\ a newspaper editorial) at $T$ discrete, equally spaced times $t=1, 2, \dots, T$. The output is modeled as the outcome of a coin toss, $S(t)$, which equals heads or tails. Thus $S(t)$ has the following properties.
\begin{enumerate}
    \item It is binary. This represents a coarse-grained approximation. In reality, an output like a newspaper editorial can be categorized along many independent dimensions (tone, policy content, and so on) and can occupy intermediate positions along a left-right spectrum in each dimension. 
    \item It is probabilistic. The signal is sampled at every $t$ from a Bernoulli distribution, $S(t) \sim \bern$, where $\theta_0$ is the true, intrinsic bias. The outcome of a coin toss is heads with probability $\bias$ and tails with probability $1-\bias$ and outcomes are not correlated temporally. Something similar occurs in reality: a media organization may deal with the complexity of current events by leaning left on day $t$ and right on day $t+1$, depending on the specific social and economic issues at play, but may harbor an internal agenda, which causes its outputs to lean left more than right (or vice versa) on average, when assessed over the long term. 
    \item It is global and public. All agents in the network simultaneously observe the coin toss and agree on the outcome. This approximation holds well for modern broadcast media (print or electronic) but breaks down for narrowcast media (e.g.\ subscription services tailored to particular social silos) or in special situations where the communication channel or carrier service curates the content selectively before delivery \cite{southwell_roles_2007}.
\end{enumerate}
The properties above can be related to the three canonical types of political media bias defined in Ref. \cite{eberl_one_2017}; visibility, tonality, and agenda bias. Visibility and tonality bias measure the relative amount of, and favorable quality of, the coverage of one viewpoint over another, respectively, and are reflected in the value of $S(t)$. Agenda bias refers to a longer-term pattern of favoring one viewpoint over another. It is reflected in the coin toss sequence and is controlled by $\theta_0$ in this paper.
\subsection{Step one of the update rule: independent observation}
\label{subsec:step1}
Let the $i$-th agent's opinion about the coin's bias at time $t$ be represented by a PDF, $\opin{i}{\theta}{t}$. The continuous variable $\theta$ is discretized to take the $k$ values $\theta_1, \dots, \theta_k$, with $k = 21$ in Refs. \cite{low_discerning_2022,low_vacillating_2022,bu_discerning_2023} and $k=2$ in the approximation at the heart of this paper. The discretization is implemented partly for computational purposes, and partly because it aligns with empirical studies of human psychology. For example, human responses to gambling tasks show evidence for dividing a continuous parameter into $\approx 16$ discrete bins \cite{tee_quantized_2019}. Expressing opinions in terms of a PDF allows agents to hold uncertain, multimodal opinions, believing equally in $\theta_i$ and $\theta_j \neq \theta_i$ for example \cite{low_discerning_2022,low_vacillating_2022,bu_discerning_2023,acemoglu_opinion_2011,parsegov_novel_2017,fang_opinion_2020}. \\

In the first step of the update rule, the $i$-th agent observes $S(t)$ and updates $\opin{i}{\theta}{t}$ accordingly by applying Bayes's rule with a Bernoulli likelihood as appropriate for a coin. The updated posterior for the $i$-th agent is given by
\begin{equation}
    \label{bayes}
    \opin{i}{\theta}{t+1/2} = \frac{\lht{t}{\theta}\opin{i}{\theta}{t}}{\sum_{\theta}\lht{t}{\theta}\opin{i}{\theta}{t}}
\end{equation}
with
\begin{equation}
    \label{likeli}
    \lht{t}{\theta} = \begin{cases}
        \theta & \text{if} \,\, S(t) \,\, \text{is heads} \\
        1-\theta & \text{if} \,\, S(t) \,\, \text{is tails.}
    \end{cases}
\end{equation}
That is, $\lht{t}{\theta}=B(\theta)$, a Bernoulli distribution with bias $\theta$. All agents share the same Bernoulli likelihood, i.e.\ they share the same internal model of the coin. The latter assumption does not apply in real-life media bias applications in general. For example, different readers of a newspaper editorial respond psychologically to its tone and content in different ways and assign different probabilities to possible values of the inferred political bias as a result. However, this simplified model enables us to analytically and empirically study conditions under which consensus and learning can be achieved. Section \ref{sec:conclusion} presents a fuller discussion of the idealizations in the model in the context of the social science literature.
\subsection{Step two of the update rule: peer pressure}
\label{subsec:step2}
The second half of the update rule captures the sharing of posterior opinions in keeping with an agent's political relationships \cite{low_discerning_2022,low_vacillating_2022,bu_discerning_2023}. Let the agents and their political relationships be indexed by a set of $n$ nodes, $\nodes$, and edges, $\edges$, respectively, which together form a connected, undirected network $G=\{\nodes,\edges\}$. The relationships are encoded by the adjacency matrix
\begin{equation}
    A_{ij} = \begin{cases}
        +1 & \text{if agents} \,\, i \,\, \text{and} \,\, j \,\, \text{communicate as allies} \\
        0 & \text{if agents} \,\, i \,\, \text{and} \,\, j \,\, \text{do not communicate} \\
        -1 & \text{if agents} \,\, i \,\, \text{and} \,\, j \,\, \text{communicate as opponents}
    \end{cases}
\end{equation}
with $A_{ij}=A_{ji}$. Agent $i$ updates their opinion, with some learning rate $\mu$, by adopting a fraction $\mu$ of the difference between their posterior opinion and the average posterior opinion of neighboring agents. Symbolically we write
\begin{equation}
    \label{fullup}
    \opin{i}{\theta}{t+1} \propto \max\left[0,\opin{i}{\theta}{t+1/2}+\mu\Delta\opin{i}{\theta}{t+1/2}\right],
\end{equation}
with
\begin{align}
    \Delta\opin{i}{\theta}{t+1/2} &= \frac{1}{d_i} \sum_{j=1}^n A_{ij}\left[\opin{j}{\theta}{t+1/2}-\opin{i}{\theta}{t+1/2}\right] \\\label{drift}
    &=-\frac{1}{d_i}\sum_{j=1}^n L_{ij} \opin{j}{\theta}{t+1/2}.
\end{align}
where $d_i=\sum_{j=1}^n \abs{A_{ij}}$ is the degree or number of agents who communicate with agent $i$. In \eqref{drift},
\begin{align}
    \label{lapl}
    L_{ij} &= s_i\delta_{ij} - A_{ij} \\\label{lapl2}
    &= \begin{cases}
        s_i & \text{if} \,\, i = j \\
        -1 & \text{if agents} \,\, i \,\, \text{and} \,\, j \neq i \,\, \text{communicate as allies} \\
        0 & \text{if agents} \,\, i \,\, \text{and} \,\, j \neq i \,\, \text{do not communicate} \\
        +1 & \text{if agents} \,\, i \,\, \text{and} \,\, j \neq i \,\, \text{communicate as opponents}
        \end{cases}
\end{align}
is the signed graph Laplacian\footnote{On an allies-only lattice, $-L_{ij}$ is the matrix obtained by discretizing the continuous Laplacian or diffusion operator, $\vb*{\nabla}^2$ \cite{hoffman_numerical_2001}, and acts as a discrete Laplacian on general allies-only networks \cite{shuman_emerging_2013}.} \cite{shi_dynamics_2018}, $s_i=\sum_{j=1}^n \mathrm{sgn}(A_{ij})$ is the number of allies minus the number of opponents who communicate with agent $i$, and $\delta_{ij}$ equals one for $i=j$ and zero otherwise. In the allies-only networks studied in this paper, $L_{ij}$ is the traditional graph Laplacian, and we have $s_i = d_i = \sum_{j=1}^n A_{ij}$, which is the degree or number of agents that communicate with agent $i$ \cite{newman_networks_2018}. The self-interaction terms in the right hand-sides of \eqref{drift} and \eqref{lapl} vanish, as we have $A_{ij}=0$ for $i=j$. The network is static for all $t$ in this paper. \\

The update step in \eqref{fullup} moves the $i$-th agent's PDF closer to (further from) the average PDF of their allies (opponents) uniformly across the full $\theta$ domain. The learning rate is constrained by $0 < \mu \leq 1/2$, so that if agent $i$ has greater belief in some $\theta$ than an allied agent $j$, their belief remains greater after the interaction\footnote{Mathematically, $\opin{i}{\theta}{t+1/2}>\opin{j}{\theta}{t+1/2}$ implies $\opin{i}{\theta}{t+1}>\opin{j}{\theta}{t+1}$ for $0<\mu\leq 1/2$. This follows by subtracting the network update step \eqref{fullup} for agent $j$ from equation \eqref{fullup} for agent $i$ to give
\begin{equation}
    \label{mu}
    \opin{i}{\theta}{t+1}-\opin{j}{\theta}{t+1} = (1-2\mu)[\opin{i}{\theta}{t+1/2}-\opin{j}{\theta}{t+1/2}],
\end{equation}
which is positive for $0<\mu\leq1/2$ assuming $\opin{i}{\theta}{t+1/2}-\opin{j}{\theta}{t+1/2}>0$.} \cite{low_discerning_2022}. This property applies specifically to two
allied agents interacting with each other but not with anybody else. The property does not apply in general
to three or more interacting allies, nor does it apply to two or more interacting opponents. The maximization operator in \eqref{fullup} ensures positivity of the PDF in the presence of opponents ($A_{ij}=-1$). In an allies-only network, the maximum and normalization in $\eqref{fullup}$ are redundant, as the matrix $\delta_{ij}-\mu L_{ij}/d_i$ is row-stochastic ($\sum_{j=1}^n \delta_{ij}-\mu L_{ij}/d_i = 1$ for all $i$ and for all $\mu$\footnote{This follows from
\begin{align}
    \label{1c}
    \sum_j \delta_{ij}-\mu L_{ij}/d_i &= 1 - \frac{\mu}{d_i}\left(d_i - \sum_j A_{ij} \right) \\\label{1f}
    &= 1
\end{align}
where we use $L_{ij}=d_i\delta_{ij}-A_{ij}$ and $d_i = \sum_j A_{ij}$ to go from \eqref{1c} to \eqref{1f}. That is, the matrix is row-stochastic for all $\mu$.}.) and positive.
\subsection{Convergence in an allies-only network without partisans}
\label{subsec:conv}
Allies-only networks without partisans that obey the two-stage update rule \eqref{bayes}--\eqref{lapl2} display three types of convergent behavior in previous studies \cite{low_discerning_2022,low_vacillating_2022,bu_discerning_2023}. First, agents converge on a common opinion or consensus, i.e. $\opin{i}{\theta}{t} \approx \opinc{t}{\theta}$ for all $i\in\nodes$, where $\opinc{t}{\theta}$ is the consensus opinion. This occurs empirically for $t \gtrsim 10^2$ for allies-only networks of all sizes \cite{low_discerning_2022}. Second, agents converge to a bimodal PDF, which is non-zero for only two $\theta$ values, i.e. $k=2$. This also occurs typically for $t\gtrsim 10^2$ \cite{low_discerning_2022}, and is consistent with behavior in other models that track multidimensional opinions \cite{demarzo_persuasion_2003}. Third, agents asymptotically learn, whereupon one obtains $\opin{i}{\theta}{t} \approx \delta(\theta-\theta')$ for some or all $i$, where $\delta(\theta-\theta')$ diverges for $\theta=\theta'$ and vanishes otherwise. For allies-only networks, this occurs typically for $t\gtrsim 10^3$ and $\theta'=\bias$, i.e.\ the agents succeed ultimately in inferring the media bias correctly.
\subsection{Disruption by partisans of convergence in an allies-only network}
\label{subsec:partisans}
The convergent behavior above in allies-only networks is disrupted by partisans \cite{bu_discerning_2023}. Partisans are obdurate agents who refuse consciously or subconsciously to be persuaded by both the media signal and the other agents. Mathematically, this corresponds to a subset of nodes, $\pnodes\subset \nodes$, which obey $\opin{i}{\theta}{t}=\opin{i}{\theta}{0}$ for all $t$ and $i \in \pnodes$.  In what follows, we make the additional, simplifying assumption, that partisan $i$ believes wholly in the truth of one $\theta$ value, $\pthetai{i}$, with $\opin{i}{\theta}{0}=\delta(\theta-\pthetai{i})$. In complete allies-only networks where all partisans agree ($\pthetai{i}=\ptheta$ for all $i\in\pnodes$), non-partisan or persuadable agents ($i \in \rnodes = \nodes\, \backslash\, \pnodes$) still converge to a bimodal consensus for $t\gtrsim 10^2$. However, consensus is not the same as asymptotic learning. Even a single partisan in a complete network can cause the consensus to vacillate between agreeing with the partisan, i.e.\ $\opinc{t}{\theta}\approx \delta(\theta-\ptheta)$, and fluctuating stochastically \cite{bu_discerning_2023}, thereby disrupting asymptotic learning. As persuadable agents still converge to a bimodal consensus, we set $k=2$ in the update rule \eqref{bayes}--\eqref{lapl2} in Sections \ref{sec:two-state} and \ref{sec:stability}. The two-state approximation makes it possible to derive an analytic instability condition for disruption by partisans (see Sections \ref{sec:two-state} and \ref{sec:stability}), whose accuracy is verified with Monte Carlo simulations in Section \ref{sec:montecarlo}.


\section{Two-state approximation for an allies-only network}
\label{sec:two-state}
The fast convergence of persuadable agents to a bimodal PDF in an allies-only network, discussed in Section \ref{subsec:partisans}, means that the disruption of asymptotic learning by partisans can be analyzed in a two-state $(k=2)$ framework to a good approximation. The two-state approximation leads to an informative analytic condition for when disruption occurs, and how the disruption condition depends on network properties, an important benefit. We approximate the update rule \eqref{bayes}--\eqref{lapl2} from Sections \ref{subsec:step1} and \ref{subsec:step2} in terms of a two-state PDF in Section \ref{subsec:updrule}. The stationary solutions of the system are written down in Section \ref{subsec:statpoint}. The results in Sections \ref{subsec:updrule} and \ref{subsec:statpoint} set the stage for the stability analysis in Section \ref{sec:stability}, where we perturb the system about the stationary solutions in Section \ref{subsec:statpoint}.
\subsection{Reformulated update rule and two-state PDF}
\label{subsec:updrule}
In the two-state approximation, we set $k=2$ in \eqref{bayes}. The PDF for agent $i$ is then
\begin{equation}
    \label{twopdf}
    \opin{i}{\theta}{t} = \begin{cases}
        \bel{i}{t} & \text{if} \,\, \theta = \theta_1 \\
        1-\bel{i}{t} & \text{if} \,\, \theta = \theta_2
    \end{cases}
\end{equation}
where $\theta_1$ and $\theta_2$ are the two possible beliefs (states). \\

The first-step of the update rule, given by \eqref{bayes} and \eqref{likeli}, combines with \eqref{twopdf} to yield
\begin{align}
    \bel{i}{t+1/2} &= \frac{\lht{t}{\theta_1}\bel{i}{t}}{\lht{t}{\theta_1}\bel{i}{t}+\lht{t}{\theta_2}\left[1-\bel{i}{t}\right]} \\\label{comp1}
    &= \frac{\bfac{t}\bel{i}{t}}{1-\left[1-\bfac{t}\right]\bel{i}{t}},
\end{align}
where $\bfac{t}=\lht{t}{\theta_1}/\lht{t}{\theta_2}$ is the Bayes factor (odds ratio) relating $\theta_1$ and $\theta_2$. In what follows, we adopt the shorthand $\bfunc{t}{\bel{i}{t}}$ to denote the right-hand side of \eqref{comp1}. Without loss of generality, we choose $\theta_1<\theta_2$ and hence have $\bfac{t}<1$, if the coin toss is heads, as $\theta_2$ is favored over $\theta_1$, and $\bfac{t}>1$ otherwise. \\

The second stage of the update rule, given by \eqref{fullup}--\eqref{lapl2}, depends on the displacement at $\theta=\theta_1$, viz.\
\begin{align}
    \Delta \opin{i}{\theta_1}{t+1/2} &= -\frac{1}{d_i}\sum_{j \in \rnodes} L_{ij}\bel{j}{t+1/2}+\frac{1}{d_i}\sum_{j \in \pnodes} A_{ij}\bel{j}{t+1/2}\\\label{beldis}
    &= -\frac{1}{d_i}\sum_{j \in \rnodes} L_{ij}\pi_j(t+1/2)+ \frac{s_{i,1}}{d_i},
\end{align}
where $s_{i,1}$ is the number of partisan allies minus the number of partisan opponents adjacent to agent $i$ with $\opin{i}{\theta}{t}=\delta(\theta-\theta_1)$ for $i \in \pnodes$. In the allies-only networks studied in this paper, we have $s_{i,1}=d_{i,1}$, where $d_{i,1}$ is the number of partisans adjacent to agent $i$ with $\opin{i}{\theta}{t}=\delta(\theta-\theta_1)$ for $i \in \pnodes$; see also the definitions following equations \eqref{lapl} and \eqref{lapl2}. There is no explicit dependence on $s_{i,2}$ in \eqref{beldis} (an implicit dependence exists through $d_i$), as partisans with $\opin{i}{\theta}{t}=\delta(\theta-\theta_2)$ have $\opin{i}{\theta_1}{t}=0$. Equation \eqref{fullup} then reduces to
\begin{equation}
    \label{eomgen}
    \bel{i}{t+1} \propto  \max\Bigg\{0,\sum_{j\in \rnodes} W_{ij} \bfunc{t}{\bel{j}{t}} + \frac{\mu s_{i,1}}{d_i}\Bigg\}
\end{equation}
In \eqref{eomgen}, $W_{ij}$ is the submatrix of $\delta_{ij} - \mu L_{ij}/d_i$ for the persuadable agents, i.e. $\delta_{ij} - \mu L_{ij}/d_i$ for $i,j \in \rnodes$. Equation \eqref{eomgen} is a special case of the general update rule \eqref{fullup}. It follows by assuming $k=2$ and by assuming that there is a subset of partisan agents, $\pnodes \subset \mathcal{N}$, obeying $\opin{i}{\theta}{t}=\delta(\theta-\theta_\mathrm{p})$ for all $t$ and $i\in\pnodes$. \\

Equations \eqref{comp1} and \eqref{eomgen} show that assuming a two-state PDF converts \eqref{bayes}--\eqref{lapl2} into a system of coupled, nonlinear, stochastic difference equations, where the opinion of each persuadable agent is described by the dynamical variable $\bel{i}{t}$. Algebraically speaking, the conversion applies equally to a mixed network containing allies and opponents, although the motivating assumption of bimodality does not hold always in a mixed network. Nevertheless, in the presence of opponents, there are three sources of nonlinearity in \eqref{comp1} and \eqref{eomgen}: one due to normalization following Bayes's rule, which gives the denominator of \eqref{comp1}, another from normalization due to the proportionality in \eqref{eomgen}, and a threshold nonlinearity due to the maximization in \eqref{eomgen}. It is unclear how to proceed analytically in the face of these three nonlinearities at the time of writing. In contrast, in an allies-only network, the update rule maintains positivity and normalization of the PDF, and \eqref{eomgen} reduces to
\begin{equation}
    \label{eom}
    \bel{i}{t+1} = \sum_{j\in \rnodes} W_{ij} \bfunc{t}{\bel{j}{t}} + \frac{\mu d_{i,1}}{d_i}.
\end{equation}
Equation \eqref{eom}, in contrast to \eqref{eomgen}, only features the nonlinearity from $\bfunc{t}{\bel{i}{t}}$, which is convex for a heads toss [$\bfac{t}<1$] and concave for a tails toss [$\bfac{t}>1$]. We analyze allies-only networks in the rest of the paper.
\subsection{Asymptotic learning, stationary solutions, and consensus}
\label{subsec:statpoint}
The evolution of a persuadable agent's PDF under the influence of partisans as $t\rightarrow\infty$ is the primary concern of this paper. If $\bel{i}{t}$ does not change through time and converges to a fixed value as $t\rightarrow\infty$, then persuadable agent $i$ is said to achieve asymptotic learning, whether or not $\bel{i}{t\rightarrow\infty}$ is the true bias or agrees with the beliefs of other agents. \\

If all persuadable agents achieve asymptotic learning, then so does the network, even if the agents' opinions disagree (i.e.\ there is no consensus), and even if some opinions are wrong. Mathematically, asymptotic learning is characterized by two properties: $\bel{i}{t}$ does not change when iterating \eqref{eom}, corresponding to a stationary solution; and small perturbations away from a stationary solution decay with time, so that the stationary solution is stable. If a stationary solution is stable then agent $i$ learns it asymptotically as $t\rightarrow\infty$. If a stationary solution is unstable, the corresponding agent typically fluctuates away from the stationary solution, exhibiting the phenomenon of turbulent nonconvergence observed in previous studies \cite{low_discerning_2022,low_vacillating_2022,bu_discerning_2023}. The mathematical definition of instability, and the conditions under which it occurs, are the subject of Section \ref{sec:stability}. Asymptotic learning for both an agent and a network refers to convergence to a stationary solution; it does not refer to the correctness or otherwise of the learned beliefs, nor does it refer to the existence or otherwise of consensus between agents. The analytic instability condition presented in Section \ref{subsec:false} refers to the disruption of a network-wide stationary solution, with $\pi_i(t\rightarrow\infty) = {\hat\pi}_i = {\rm constant}$ for all $i$ before the stationary solution is perturbed. \\

Stationary solutions satisfy $\bel{i}{t+1}=\bel{i}{t}=\stat_i$. They fall into two classes: trivial solutions, namely $\stat_i=0$ or $\stat_i=1$, and non-trivial solutions, which are complicated functions of $\mu$, $\bfac{t}$, and an agent's connections to partisans and other persuadable agents. Interestingly, however, nontrivial solutions of $\pi_i(t+1) = \pi_i(t)$ fluctuate in response to $S(t)$, because they depend on $\Lambda(t)$. Hence they cannot be learnt asymptotically by persuadable agents. To see why, consider for example a single persuadable agent connected to a partisan with $x_{\mathrm{p}}(t,\theta)=\delta(\theta-\theta_1)$. Equation \eqref{eom} then takes the form $\pi(t+1) = (1-\mu)\bfunc{t}{\pi(t)}+\mu$, which has two stationary solutions $\hat{\pi}=1$ and $\hat{\pi} = \mu/[1-\bfac{t}]$. The first corresponds to agreement with the partisan that the coin has bias $\theta_1$ and is a solution for both possible outcomes of the coin toss at $t$. The second solution only obeys $0\leq \hat{\pi}\leq 1$, as long as the coin toss returns heads at $t$, as one has $\bfac{t} \leq 1-\mu$. While it is a stationary solution, it cannot be learnt asymptotically as it is only unchanged by the update rule, if the coin toss is heads; a tails toss gives $\pi(t+1) \neq \pi(t)$. As this paper focuses on asymptotic learning and its stability, we do not analyze the nontrivial solutions below. \\

Trivial stationary solutions correspond to consensus in the sense defined in Section \ref{subsec:partisans}, as long as the persuadable subnetwork is connected, with $\sum_{j\in \rnodes} A_{ij}\neq 0 $ for all $i \in \rnodes$. This property is consistent with empirical results published previously from Monte Carlo simulations on complete networks \cite{bu_discerning_2023}. To see why, we split the persuadable agents into two groups, $\rnodes = \mathcal{N}_{\mathrm{r},1} \cup \mathcal{N}_{\mathrm{r},2}$ and assume that each group collectively obeys a different trivial stationary solution, i.e.\ $\stat_i = 0$ for all $i \in \mathcal{N}_{\mathrm{r},2}$, and $\stat_i=1$ for all $i \in \mathcal{N}_{\mathrm{r},1}$. For an agent in $\mathcal{N}_{\mathrm{r},2}$, we have $\stat_i=0$ and hence obtain
\begin{align}
    \label{trivstat0}
    0 &= \sum_{j \in \mathcal{N}_{\mathrm{r},2}} W_{ij} \bfunc{t}{0} + \sum_{j \in \mathcal{N}_{\mathrm{r},1}} W_{ij} \bfunc{t}{1} + \frac{\mu d_{i,1}}{d_i} \\\label{trivstat1}
    &= \frac{\mu}{d_i}\sum_{j \in \mathcal{N}_{\mathrm{r},1}} A_{ij} + \frac{\mu d_{i,1}}{d_i},
\end{align}
where we use $\bfunc{t}{0}=0$ and $\bfunc{t}{1}=1$ to go from \eqref{trivstat0} to \eqref{trivstat1}, and $i \notin \mathcal{N}_{\mathrm{r},1}$ to reduce $W_{ij}$ to $\mu A_{ij}/d_i$. Equation \eqref{trivstat1} requires $\sum_{j\in\mathcal{N}_{\mathrm{r},1}} A_{ij}+d_{i,1}=0$, which means that both terms are zero as they are non-negative. Hence all agents in $\mathcal{N}_{\mathrm{r},2}$ must be disconnected from persuadable agents or partisans who disagree with $\stat_i=0$. An identical calculation for an agent in $\mathcal{N}_{\mathrm{r},1}$ gives an analogous condition, $\sum_{j\in\mathcal{N}_{\mathrm{r},2}} A_{ij}+d_{i,2}=0$. Hence the subnetworks corresponding to $\mathcal{N}_{\mathrm{r},1}$ and $\mathcal{N}_{\mathrm{r},2}$ must be disconnected from one another, and neither contain partisans who disagree with their respective trivial stationary solution. \\

When a network disconnects into two subnetworks, its dynamics decouple and can be analyzed separately. This yields three insights for asymptotic learning. First, for a connected persuadable subnetwork to support a trivial stationary solution it must also be a consensus, i.e.\ $\stat_i = \belcs$ for all $i \in \rnodes$ with either $\belcs=0$ or $\belcs=1$. Stationarity, and hence asymptotic learning, are global behaviors, not achievable as an individual without the agreement of the persuadable subnetwork. Second, for $\belcs=0$ or $\belcs=1$ to be a solution, all partisans must have $\opin{i}{\theta}{t}=\delta(\theta-\theta_2)$ or $\opin{i}{\theta}{t}=\delta(\theta-\theta_1)$, respectively. Asymptotic learning requires that all partisans agree with one another; cf.\ dueling partisans studied in Section 3.3 in Ref. \cite{bu_discerning_2023}. Third, partisans define what can be learnt asymptotically by the persuadable subnetwork. If one has (for example) $\theta_2=\bias$ and $\opin{i}{\theta}{t}=\delta(\theta-\theta_1)$ for all $i\in\pnodes$, then persuadable agents can learn the false bias but not the true bias. These latter two insights shape the condition for disruption of asymptotic learning by partisans, i.e. the presence of one partisan in the network, who believes in a false bias, means that persuadable agents cannot asymptotically learn the true bias.

\section{Analytic instability condition}
\label{sec:stability}
In this section, we assess the stability of the stationary solutions identified in Section \ref{subsec:statpoint}. In Section \ref{subsec:perturb}, we perturb the system infinitesimally away from its stationary solution and derive an analytic condition for the perturbation to grow (unstable) or decay (stable) with time. In Section \ref{subsec:true}, we apply the results from Section \ref{subsec:perturb} to show that networks with no partisans and networks with partisans with $\opin{i}{\theta}{t}=\delta(\theta-\theta_0)$ asymptotically learn the true bias of the coin. In Section \ref{subsec:false}, we again apply the results from Section \ref{subsec:perturb} to networks that are disrupted from asymptotically learning the true bias, as described in Section \ref{subsec:statpoint}. We derive an analytic condition for when partisans lure persuadable agents into converging on a false bias and when they stop asymptotic learning happening at all, so that persuadable agents experience turbulent nonconvergence \cite{bu_discerning_2023}. The former and latter types of disruption occur when the stationary consensus is stable and unstable respectively.
\subsection{Perturbations about a stationary solution}
\label{subsec:perturb}
If the stationary consensus $\bel{i}{t} = \stat_i = \belcs$ found in Section \ref{subsec:statpoint} is stable, the persuadable agents achieve asymptotic learning. In \ref{sec:appendix_a}, we perturb the system away from the stationary consensus by writing $\bel{i}{t}=\belcs+\delta_i(t)$, with $| \delta_i(t) |$ small, and linearize the update rule \eqref{eom}. This gives an update rule for the vector of perturbations $\vb*{\delta}(t)=\left[\delta_1(t),\ldots,\delta_{\abs{\rnodes}}(t)\right]^\top$. If the fluctuation amplitude, $\norm{\vb*{\delta}(t)}$, grows in the limit $t\rightarrow\infty$, the stationary consensus is unstable, where $\norm{\cdot}$ is a standard vector norm. The limit $t \rightarrow \infty$ is taken probabilistically; formally speaking, one says that the system is unstable with probability one. This means that some pathological, finite coin toss sequences contradict the instability result, but when the ensemble of all sequences is considered, the probability of a pathological finite sequence being selected tends to zero as $t\rightarrow\infty$\footnote{This definition of stability is called almost sure stability in Ref. \cite{shaikhet_lyapunov_2011}.}. For $\belcs=1$, \ref{sec:appendix_a} shows that the stationary consensus is unstable for
\begin{equation}
    \label{stabcond}
    \kldivbern{\theta_1}-\kldivbern{\theta_2} > -\log \rho(\vb{W}).
\end{equation}
The instability condition for $\belcs=0$ is the same as \eqref{stabcond} with $\theta_1$ and $\theta_2$ swapped everywhere. In \eqref{stabcond}, $\rho(\vb{W})=\max_i\lambda_i(\vb{W})$ is the spectral radius or largest eigenvalue of the matrix $\vb{W}$ introduced in \eqref{eomgen}, and $\mathrm{KL}$ denotes the Kullback-Leibler divergence \cite{lehmann_testing_2005} (sometimes called relative entropy \cite{cover_elements_2005}), given by
\begin{equation}
    \label{kldiv}
        \kldivbern{\theta}=\sum_{S(t) \in \{0,1\}} \bern \log\Bigg\{\frac{\bern}{\lht{t}{\theta}}\Bigg\}.
\end{equation}
The sum in \eqref{kldiv} is over heads [$S(t)=1$] and tails [$S(t)=0$]. The KL divergence measures the similarity between two PDFs, here $\bern$ and $\lht{t}{\theta}$\footnote{See the Supplementary Information in Ref. \cite{machta_parameter_2013} for an intuitive, information theoretic introduction to the KL divergence.}. Given a coin toss sequence sampled from $\bern$, equation \eqref{kldiv} quantifies, as the sequence becomes infinitely long, how confidently an agent can distinguish that the sequence is sampled from $\bern$ and not $\lht{t}{\theta}$. If the confidence is high, then \eqref{kldiv} is large; otherwise, equation \eqref{kldiv} approaches zero. Therefore, the left-hand side of \eqref{stabcond} is negative, or positive, if the bias is closer to $\theta_1$, or $\theta_2$, respectively. \\

A popular method of statistical hypothesis testing is the likelihood ratio test, where one hypothesis is selected over another if the ratio of likelihoods exceeds a threshold \cite{lehmann_testing_2005}. When a data set is large, and its elements are statistically independent, the stability condition [negation of \eqref{stabcond}] is equivalent to the likelihood ratio test, with $-\log\rho(\vb{W})$ being the threshold criterion for when the hypothesis $\bias=\theta_1$ is selected over $\bias=\theta_2$ \cite{lehmann_testing_2005,eguchi_interpreting_2006}. The Neyman-Pearson lemma states that the likelihood ratio test is optimal in that it minimizes the probability of false negatives (rejecting null hypotheses while they are true) subject to a constraint on the probability of false positives (accepting alternative hypotheses while they are false) \cite{lehmann_testing_2005}.
\subsection{Stable consensus about a true bias}
\label{subsec:true}
One implication of \eqref{stabcond} is that, in allies-only networks that do not contain partisans, asymptotic learning is always achieved. This follows from $\vb{W}$ being row-stochastic, which implies $\rho(\vb{W})=1$ \cite{meyer_matrix_2023}, so that the right-hand side of $\eqref{stabcond}$ vanishes. The stationary consensus $\belcs=1$, which corresponds to certainty that the bias equals $\theta_1$, is stable by \eqref{stabcond} for $\kldivbern{\theta_1}<\kldivbern{\theta_2}$. That is, agents asymptotically learn that the bias is $\theta_1$, if $\lht{t}{\theta_1}$ is ``closer" to $\bern$ than $\lht{t}{\theta_2}$ in the KL sense. An analogous argument can be made for $\belcs=0$. Hence an allies-only network with no partisans asymptotically learns that the coin has the bias which minimizes the KL-divergence with $\bern$. In what follows, we fix $\theta_2=\bias$, which means that networks without partisans asymptotically learn the true bias, because one has $\kldivbern{\theta_2}=0$ and $\kldivbern{\theta_1}>0$. \\

\begin{figure}[H]%
    \centering
    \includegraphics[width=\linewidth]{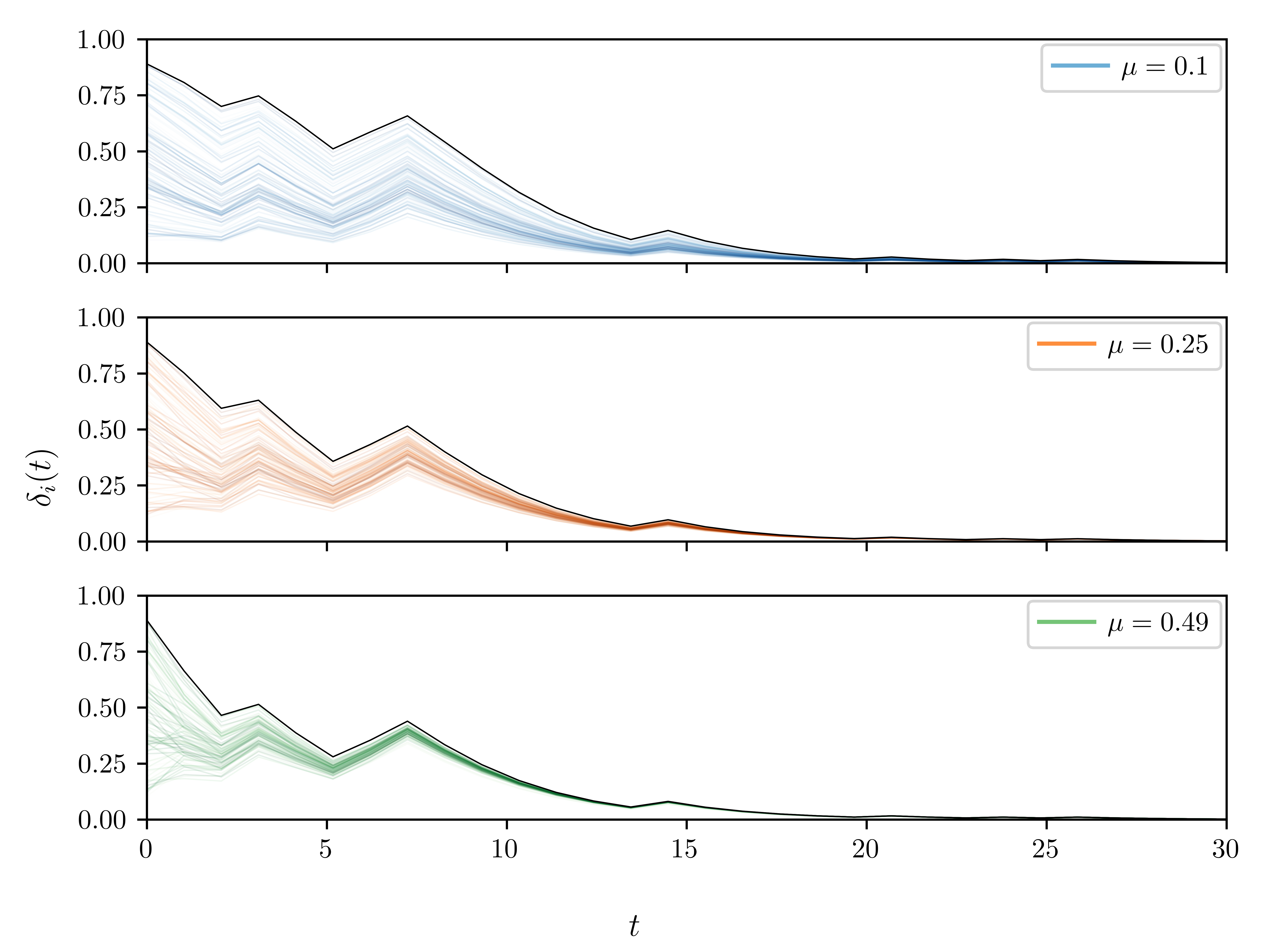}
    \caption{Stable consensus about the true bias: evolution of $\abs{\delta_i(t)}$ about $\belcs=0$ for a Barab\'asi-Albert (BA) network \cite{barabasi_emergence_1999} with $n=100$, $m=5$, $\abs{\pnodes}/n=0.01$, for a coin toss sequence of length $T=30$ with $\theta_1=1-\bias$ and $\theta_2 = \bias = 0.6$. The coin toss sequence, priors (how priors are selected is discussed in Section \ref{subsec:control}), and network are fixed for three simulations with learning rates $\mu=0.1 \, \text{(top panel; blue curves)},0.25 \, \text{(middle panel; orange curves)},0.49 \, \text{(bottom panel; green curves)}$. There are $\abs{\rnodes}$ colored curves in each panel; each curve corresponds to $\abs{\delta_i(t)}$ for some $i \in \rnodes$. The black curve corresponds to the maximum fluctuation amplitude, $\norm{\vb*{\delta}(t)}=\max_i \abs{\delta_i(t)}$.}
    \label{fig:dyn_stable}
\end{figure}

Next, we consider an allies-only network containing partisans, who believe with certainty that the bias is the true bias. Upon choosing $\theta_2 = \bias$ without loss of generality, one concludes that only $\belcs=0$ can be learnt asymptotically (see Section \ref{subsec:statpoint}). The persuadable agents are then disrupted from learning the true bias if one has
\begin{equation}
    \label{stabcond3}
    \kldivbern{\theta_1} < \log \rho(\vb{W}).
\end{equation}
As $\vb{W}$ is a submatrix of a row-stochastic matrix, it is row-substochastic ($\sum_jW_{ij}<1$) and obeys $\rho(\vb{W})<1$ \cite{meyer_matrix_2023}. Hence the instability condition \eqref{stabcond3} is not satisfied, because the KL-divergence is non-negative. If the partisans believe with certainty in the true bias, then the persuadable agents are guaranteed to agree with them asymptotically. This result holds independent of all model parameters. An example of this, for a representative network and coin toss sequence, is displayed in Figure \ref{fig:dyn_stable}, where we graph the amplitude $|\delta_i(t)|$ of the perturbation away from the stationary consensus $\belcs = 0$ for every persuadable agent $i \in \rnodes$ as a function of time $t$. The black curve is the agent with the greatest disagreement with the consensus as measured by $\norm{\vb*{\delta}(t)}=\max_{i\in\rnodes}\abs{\delta_i(t)}$, as discussed in \ref{sec:appendix_a}. For $\mu=0.1,0.25,0.49$, persuadable agents quickly reach consensus for $t \gtrsim 10$ slightly before asymptotically learning the true bias, with $\norm{\vb*{\delta}(t)}$ bounding every trajectory from above.
\subsection{Disrupted consensus about a false bias and turbulent nonconvergence}
\label{subsec:false}
\begin{figure}[H]%
    \centering
    \includegraphics[width=\linewidth]{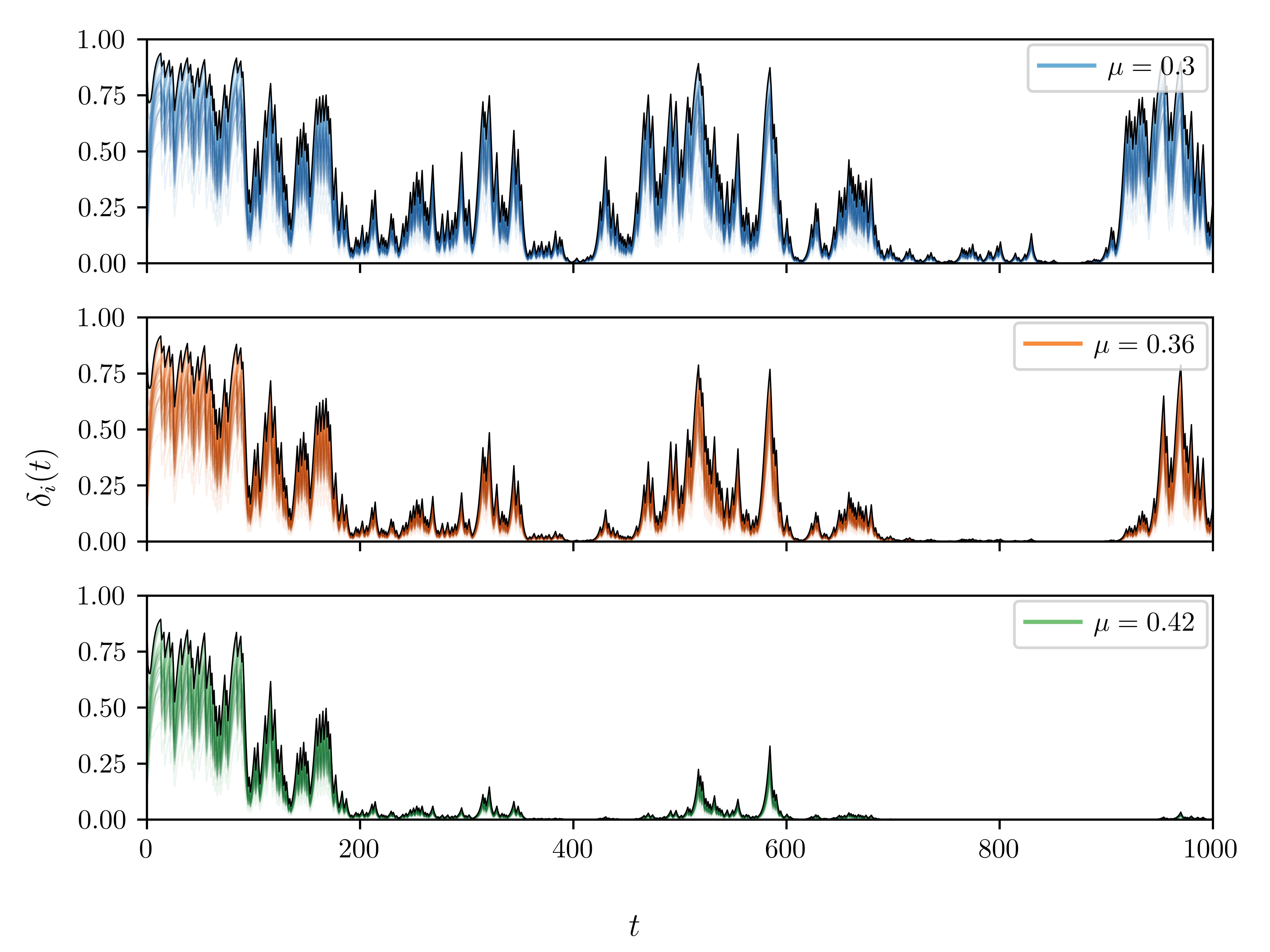}
    \caption{As for Figure \ref{fig:dyn_stable}, but for disrupted consensus about a false bias. Learning rates, partisan fraction, and simulation length are changed to $\mu=0.3,0.36,0.42$, $\abs{\pnodes}/n=0.2$, and $T=10^3$ to respectively exemplify instability (top panel; blue curves), marginal (in)stability boundary (middle panel; orange curves), and stability (bottom panel; green curves) behavior, as demarcated by \eqref{stabcond2}.}
    \label{fig:dyn_unstable}
\end{figure}

Now suppose that an allies-only network contains partisans, who believe with certainty that the bias is not the true bias. Under these conditions, the partisans disrupt the persuadable agents, who can only learn asymptotically the false bias through peer pressure (see Section \ref{subsec:statpoint}) but receive a contradictory statistical signal from the coin tosses. The competing stimuli produce a network-mediated form of cognitive dissonance. The dynamics can be stable or unstable, depending on the parameter values. Upon setting $\theta_2 = \theta_0$ without loss of generality, and perturbing the false consensus $\belcs=1$, equation \eqref{stabcond} implies the instability condition
\begin{equation}
    \label{stabcond2}
    \kldivbern{\theta_1} > -\log \rho(\vb{W}).
\end{equation}
Both sides of \eqref{stabcond2} are non-negative, and stability and instability are both possible. An example of the dynamics in the unstable and stable regimes, for a representative network and coin toss sequence, can be seen in Figure \ref{fig:dyn_unstable}, where we graph the perturbation amplitudes $\abs{\delta_i(t)}$ away from $\belcs=1$ versus time as in Figure \ref{fig:dyn_stable}. We choose $\mu = 0.3$ (blue curves), $0.36$ (orange curves), $0.42$ (green curves) to display the unstable, marginally stable, and stable regimes, respectively. In the unstable regime, in the top panel of Figure \ref{fig:dyn_unstable}, $\kldivbern{\theta_1} > -\log \rho(\vb{W})$, agents vacillate intermittently between extended intervals (covering multiple time steps) during which they agree with the partisans and extended intervals when their PDFs fluctuate stochastically. The fluctuations, termed turbulent nonconvergence, are also observed in the systematic Monte Carlo simulations performed previously \cite{low_discerning_2022,bu_discerning_2023}. The persuadable agents reach consensus for $t\gtrsim 200$ as described in Section \ref{subsec:partisans} before exhibiting turbulent nonconvergence in unison for the remaining time steps. There is a brief period of agreement with the partisans for $800 \lesssim t \lesssim 900$ before returning to turbulent nonconvergence. The marginally stable regime, $\kldivbern{\theta_1} = -\log \rho(\vb{W})$, can be viewed in the middle panel of Figure \ref{fig:dyn_unstable}. The amplitude switching intermittently between agreeing with the partisans and turbulent convergence. The onset of turbulent nonconvergence happens similarly in the top and middle panels. In the stable regime, $\kldivbern{\theta_1} < -\log \rho(\vb{W})$, the partisan influence is strong enough to ultimately convince the agents of the false bias despite the coin advising them otherwise. This can be seen in the bottom panel of Figure \ref{fig:dyn_unstable} where the persuadable agents reach consensus for $t\gtrsim 100$ before asymptotically learning the false bias for $t \gtrsim 200$.
In contrast to \eqref{stabcond3}, the instability condition \eqref{stabcond2} is sensitively dependent on the control parameters and demarcates where the exogenous influence of the coin wins out over the endogenous peer pressure from and causes turbulent nonconvergence. We verify \eqref{stabcond2} numerically in Section \ref{sec:montecarlo}.

\section{Monte Carlo simulations}
\label{sec:montecarlo}
In this section, we verify the analytic instability condition \eqref{stabcond2} in Section \ref{subsec:false} and the two associated modes of partisan disruption with Monte Carlo multi-agent simulations\footnote{The simulation code is shared publicly at \url{https://github.com/JarraHorstman/Opinion-dynamics}}. In Section \ref{subsec:control}, we discuss how the control parameters unrelated to the network ($\mu$, $k$, $\theta_1$, $\bias$, coin toss sequences, and initial priors) are varied and explain how we classify whether a simulation achieves asymptotic learning or not. In Section \ref{subsec:size}, we study the effect of the network's size by simulating networks with $10 \leq n \leq 10^2$ in steps of $10$ while keeping the attachment parameter $m$ and partisan fraction $\abs{\pnodes}/n$ fixed. In Section \ref{subsec:spars}, we study the effect of sparsity by varying $m$ over the range $5 \leq m \leq 15$ in steps of one while keeping $n$ and $\abs{\pnodes}/n$ fixed. In Section \ref{subsec:frac}, we study the effect of partisan fraction by testing $10$ values over the range $0.01 \leq \abs{\pnodes}/n \leq 0.3$ while keeping $n$ and $m$ fixed. For each choice of $n$, $m$, and $\abs{\pnodes}/n$, we generate $10^2$ different networks to capture statistical fluctuations in the preferential attachment mechanism and partisan placement. Table \ref{table1} summarizes the parameter settings in the numerical experiments in Sections \ref{subsec:size}--\ref{subsec:frac}, along with the analogous settings in Sections \ref{subsec:true} and \ref{subsec:false} for ease of comparison.
\begin{table}[H]%
    \centering
    \begin{tabular}{lccccc}
    \hline
                           & $\theta_{\mathrm{p}}$ & $n$ & $m$ & $|\mathcal{N}_{\rm{p}}|/n$ & $\mu$ \\ \hline
    Sec. 4.2 & $\theta_0$ & $10^2$ & $5$ & $0.01$ & $[0.10,0.49]$ \\
    Sec. 4.3 & $1-\theta_0$ & $10^2$ & $5$ & $0.2$ & $[0.30,0.42]$ \\
    Sec. 5.2 & $1-\theta_0$ & $[10,10^2]$ & $5$ & $0.2$ & $[0.10,0.49]$ \\
    Sec. 5.3 & $1-\theta_0$ & $10^2$ & $[5,15]$ & $0.2$ & $[0.10,0.49]$ \\
    Sec. 5.4 & $1-\theta_0$ & $10^2$ & $5$ & $[0.01,0.30]$ & $[0.10,0.49]$  \\ \hline
    \end{tabular}
    \caption{Parameter settings for the numerical experiments performed in this paper, together with the sections in which the experiments are discussed. A single value indicates that the parameter is fixed. A range of values in the format $[a,b]$ indicates that the parameter varies from $a$ to $b$. Experiments are distinguished by whether the partisan believes in the true bias ($\theta_{\mathrm{p}} = \theta_0$) or a false bias ($\theta_{\mathrm{p}} \neq \theta_0$).}
    \label{table1}
\end{table}
\subsection{Control parameters and simulation set-up}
\label{subsec:control}
To verify the instability condition \eqref{stabcond2} with Monte Carlo simulations, we need an objective criterion for deciding whether the output data $\bel{i}{t}$ for all $i\in\rnodes$ generated by any particular simulation correspond to asymptotic learning or not. To this end, we run simulations for $10^3$ different coin toss sequences, each of length $T=10^3$, for a given network. The simulations are computed with the two-state ($k=2$) approximation with belief PDFs updated according to the reformulated update rule \eqref{eom}. The empirical cumulative distribution function (CDF) for the fluctuation amplitude at the final time-step across all coin toss sequences, $\mathrm{Pr}[\norm{\vb*{\delta}(T)}\leq\varepsilon]$, is then calculated. Equation \eqref{stabcond2} is verified visually, when simulations with control parameters in the unstable regime return $\mathrm{Pr}[\norm{\vb*{\delta}(T)}\leq\varepsilon=0.05]\approx 0$, and control parameters in the stable regime return $\mathrm{Pr}[\norm{\vb*{\delta}(T)}\leq\varepsilon]\approx 1$. As discussed in \ref{sec:appendix_a}, $\norm{\cdot}$ is the $L_\infty$-norm, i.e. $\norm{\vb*{\delta}(T)} = \max_i \abs{\delta_i(T)}$, and measures the largest disagreement with the stationary consensus for any agent in the persuadable subnetwork at $t=T$. Priors are selected for each simulation by randomly generating $0.1 \leq \bel{i}{0} \leq 0.9$ for all $i \in \rnodes$. We fix $\theta_1=1-\bias$ and $\theta_2 = \bias=0.6$ and take $10$ learning rate values between $0.1 \leq \mu \leq 0.49$ for each network. \\

One key goal of the Monte Carlo tests in this section is to verify and explore the instability condition \eqref{stabcond2} for different kinds of networks. The network properties enter \eqref{stabcond2} through $\rho(\vb{W})=1-\mu \plam$, where $\plam$ is the smallest eigenvalue of the persuadable agent submatrix of $L_{ij}/d_i$. The instability condition \eqref{stabcond2} is then $\kldivbern{\theta_1}>-\log(1-\mu\plam)$. The eigenvalue $\plam$ contains information about the size of, connections within, and connections between the persuadable and partisan subnetworks. Therefore, the instability condition \eqref{stabcond2} relates network properties to the learning rate. For general networks, $\plam$ is difficult to calculate analytically. However, the submatrix of $L_{ij}$ for $i,j\in \rnodes$ has been studied and is called the grounded Laplacian \cite{pirani_spectral_2014,pirani_smallest_2014,liu_optimizing_2021}. There is no general expression for the smallest eigenvalue of the grounded Laplacian but there are asymptotic results for large, random networks \cite{pirani_smallest_2014}. The grounded Laplacian has been investigated as a network centrality measure \cite{pirani_spectral_2014} and as a controller of networked dynamical systems \cite{liu_optimizing_2021}. \\

In this paper, we vary $\plam$ by independently varying the size, attachment parameter, and partisan fraction of Barab\'asi-Albert (BA) networks \cite{barabasi_emergence_1999}. BA networks are the archetypal model of scale-free networks, which have a power-law degree distribution. Many empirical studies of real-world social networks find them to be approximately scale-free \cite{newman_networks_2018}. A BA network is characterized by two parameters: $n$, the size of the network, and $m$, the attachment parameter, which controls sparsity. Starting from a complete graph of $m$ nodes, a new node is preferentially attached to $m$ existing nodes with high degrees. This process is repeated until the number of nodes equals $n$. For $m = n-1$, a BA network is complete with $d_i=n-1$ for all $i$. Partisans are included in the network by randomly and uniformly designating $\abs{\pnodes}$ nodes to have fixed belief PDFs.  
\subsection{Network size}
\label{subsec:size}
\begin{figure}[H]%
    \centering
    \begin{subfigure}[b]{0.49\linewidth}
        \centering
        \includegraphics[width=\linewidth]{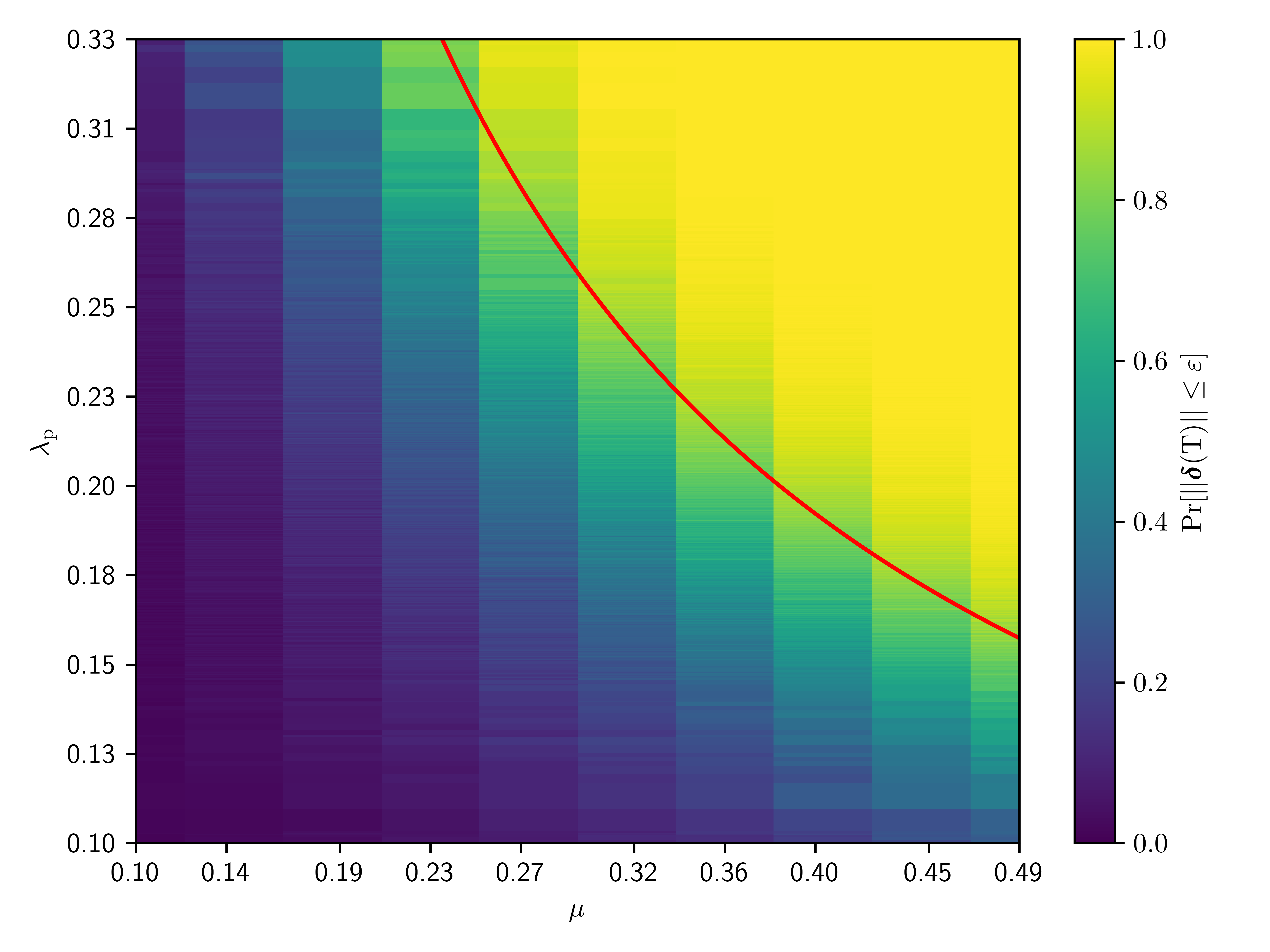}
        \caption{}
        \label{subfig:size_conv}
    \end{subfigure}
    \begin{subfigure}[b]{0.49\linewidth}
        \centering
        \includegraphics[width=\linewidth]{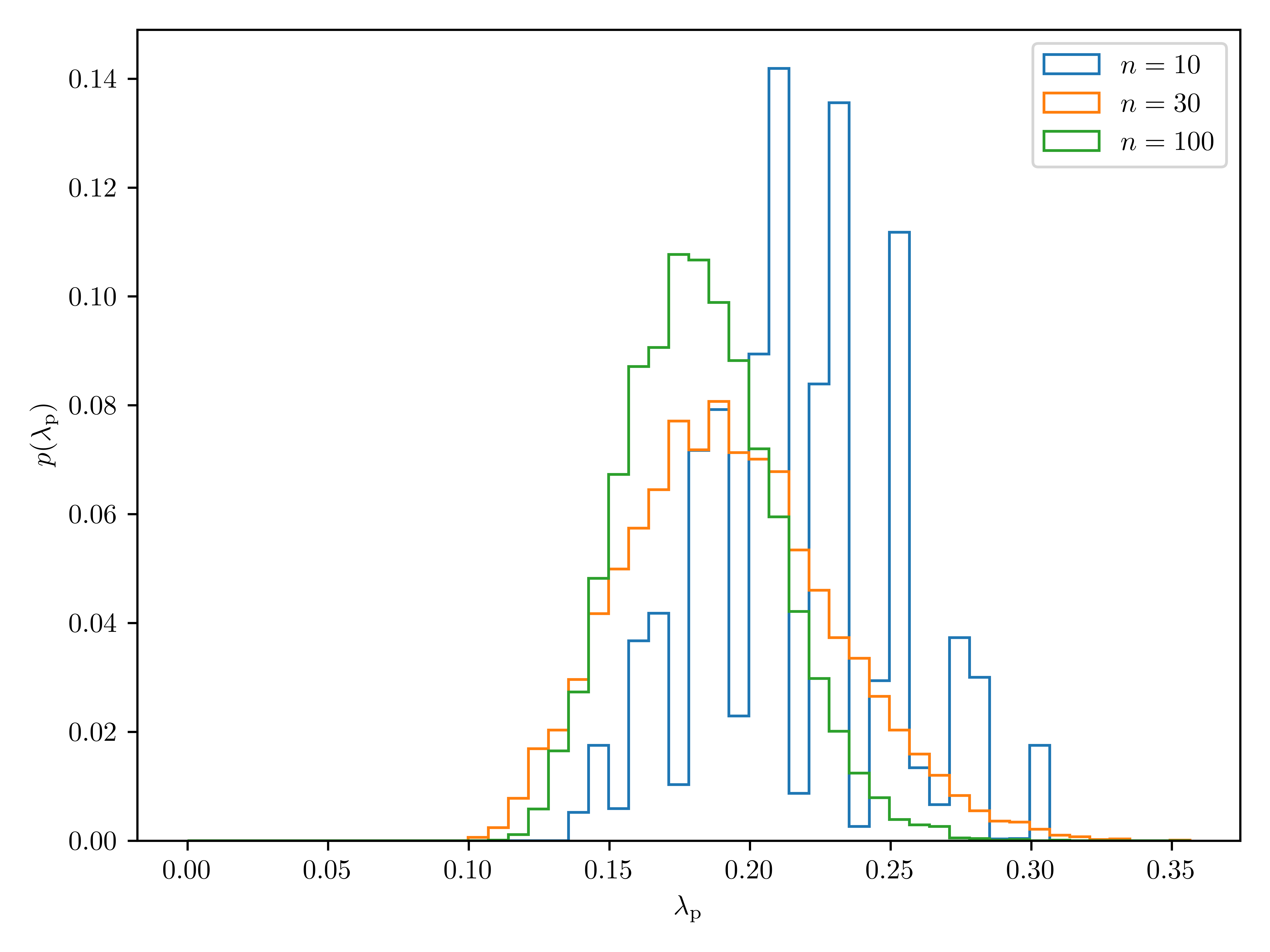}
        \caption{}
        \label{subfig:size_eval}
    \end{subfigure}
    \caption{Instability of asymptotic learning of the stationary consensus $\belcs=1$ in the two-state ($k=2$) approximation as a function of $\lambda_{\rm p}$ and $\mu$, for $10$ equally spaced network sizes in the range $10\leq n \leq 10^2$, with sparsity ($m=5$) and partisan fraction ($\abs{\pnodes}/n=0.2$) fixed. (a) Gridded heatmap of the fluctuation amplitude CDF $\Pr[\norm{\vb*{\delta}(T)} \leq \varepsilon=0.05]$, scaled according to the color bar at right. The red curve represents the analytic instability threshold from \eqref{stabcond2}, i.e.\ $\kldivbern{\theta_1}=-\log(1-\mu\plam)$. (b) Histogram of $\plam$ for $10^4$ networks at network sizes $n=10 \, \text{(blue curve)},30 \, \text{(orange curve)},10^2 \, \text{(green curve)}$. The $10^2$ networks per parameter value featured in the left-hand figure represent a subsample drawn from the $10^4$ networks in the right-hand figure.}
    \label{fig:size}
\end{figure}

In the first set of Monte Carlo tests, we vary the network size $n$ while holding $m$ and $\abs{\pnodes}/n$ fixed. Figure \ref{fig:size} displays the results. Specifically, Figure \ref{subfig:size_conv} displays the fluctuation amplitude CDF as a function of $\mu$ and $\plam$, such that $\plam$ varies, as $10 \leq n \leq 10^2$ varies, and $m=5$ and $\abs{\pnodes}/n = 0.2$ stay fixed. The left panel of Figure \ref{subfig:size_conv} graphs the probability of obtaining $\norm{\vb*{\delta}(T)} \leq \varepsilon = 0.05$ for the simulation ensemble defined in Section \ref{subsec:control}. Yellow and blue indicate high and low probability respectively. The red curve corresponds to the instability condition $\kldivbern{\theta_1}=-\log(1-\mu\plam)$, i.e.\ equation \eqref{stabcond2}. Additionally, Figure \ref{subfig:size_eval} displays the histogram of $\lambda_{\rm p}$ values for $n=10$, $30$, and $10^2$, with $10^4$ networks per $n$ value. The $10^2$ networks for each parameter choice featured in Figure \ref{subfig:size_conv} represent a subsample drawn from the $10^4$ networks in Figure \ref{subfig:size_eval}. As there are $10$ values for $n$ between $10\leq n \leq 10^2$, there are $10^3$ networks for each value of $\mu$ in Figure \ref{subfig:size_conv}. Each of these networks corresponds to one cell in Figure \ref{subfig:size_conv}, colored according to $\mathrm{Pr}[\norm{\vb*{\delta}(T)}\leq \varepsilon]$. In making this figure we use adaptive binning, so that networks with densely graduated $\lambda_{\mathrm{p}}$ values occupy thinner (in the vertical direction) rectangular cells. This can be seen near the top and bottom margins of Figure \ref{subfig:size_conv}.
Figure \ref{subfig:size_conv} confirms visually the analytic formula \eqref{stabcond2} for networks with $10\leq n \leq 10^2$. Equation \eqref{stabcond2} predicts that networks where $\plam$ lies above the red curve in Figure \ref{subfig:size_conv} should achieve asymptotic learning of $\belcs=1$ for most coin-toss sequences. The simulations agree: the grid cells above the red curve are predominantly yellow, indicating that the fluctuation amplitude approaches zero at $t=T$ for most coin toss sequences, and hence the agents asymptotically learn the false bias. Conversely, equation \eqref{stabcond2} predicts that networks with $\plam$ below the red curve experience turbulent nonconvergence. Again, the simulations agree: the grid cells below the red curve are predominantly blue, indicating that the fluctuation amplitude is nonzero at $t=T$ for most coin toss sequences, and hence the agents undergo turbulent nonconvergence away from the stationary consensus. \\

Figure \ref{subfig:size_eval} shows how the distribution of $\plam$ values sampled in Figure \ref{subfig:size_conv} depends on the network size $n$, reflecting the randomness of preferential attachment and partisan placement. For $n=30$ and $n=10^2$, $p(\plam)$ is unimodal, but it is not for $n=10$. For small networks, we find $\langle\plam\rangle>\abs{\pnodes}/n$. Specifically, we find $\langle \plam \rangle=0.21$, $0.19$, and $0.18$ for $n=10$, $30$, and $10^2$, respectively. Numerical investigation of larger networks (not plotted) return $\langle\plam\rangle=0.18$ for $10^3 \leq n \leq 10^4$. That is, $\langle\lambda_{\rm p}\rangle$ asymptotes to a constant value less than the partisan fraction, as $n$ increases. Therefore, persuadable agents in a large network, for a given coin toss sequence, are more likely to experience turbulent nonconvergence than in a small network. The standard deviation around $\langle\plam\rangle$ decreases with $n$, returning $6.4\cross 10^{-2}$, $3.7\cross 10^{-2}$, and $2.6\cross 10^{-2}$ for $n=10$, $30$, and $10^2$ respectively, and $4.4\cross 10^{-3}$ for $n=10^4$.
\subsection{Network sparsity}
\label{subsec:spars}
\begin{figure}[H]%
    \centering
    \begin{subfigure}[b]{0.49\linewidth}
        \centering
        \includegraphics[width=\linewidth]{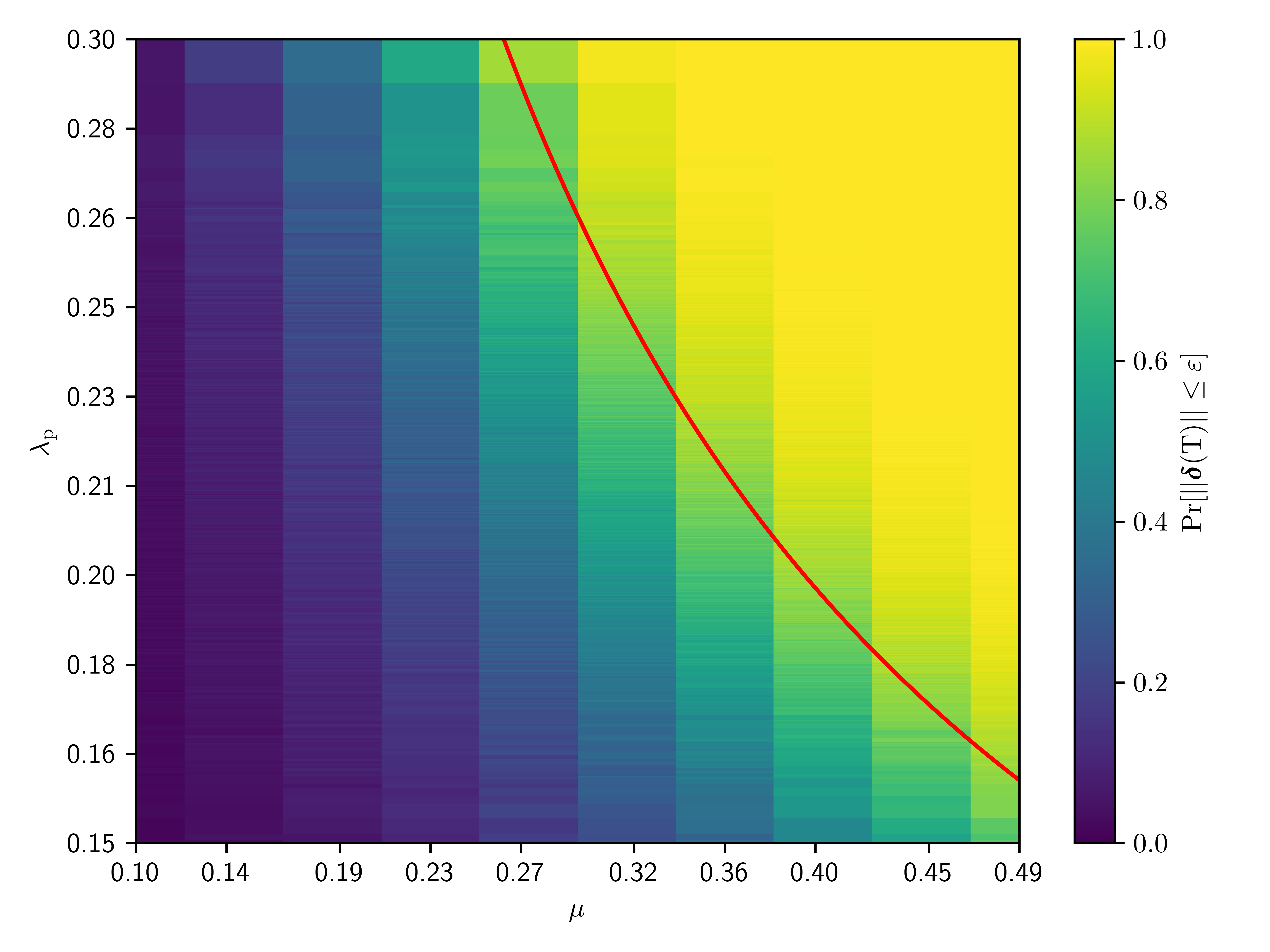}
        \caption{}
        \label{subfig:spars_conv}
    \end{subfigure}
    \begin{subfigure}[b]{0.49\linewidth}
        \centering
        \includegraphics[width=\linewidth]{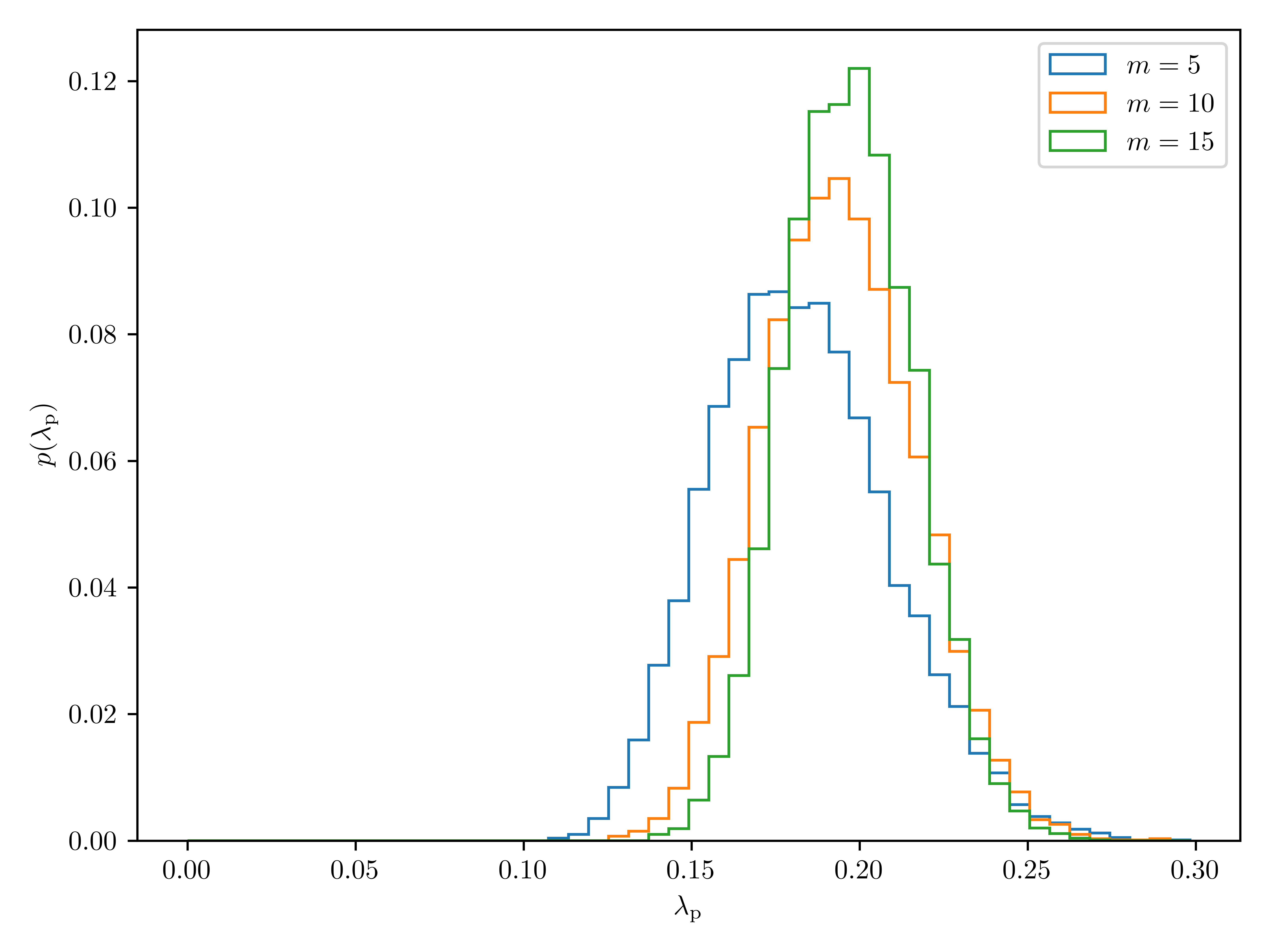}
        \caption{}
        \label{subfig:spars_eval}
    \end{subfigure}
    \caption{Effect of network sparsity: as for Figure \ref{subfig:size_conv}, but with $5 \leq m \leq 15$ and $n=10^2$ and $\abs{\pnodes}/n=0.2$ fixed. (a) Heatmap of the fluctuation amplitude CDF. The red curve corresponds to the analytic instability threshold \eqref{stabcond2}. (b) Histogram of $\plam$ for $m=5 \, \text{(blue curve)},10 \, \text{(orange curve)},15 \, \text{(green curve)}$.}
    \label{fig:spars}
\end{figure}

In the next set of Monte Carlo tests, we vary the network sparsity through the BA attachment parameter $m$ while holding $n$ and $\abs{\pnodes}/n$ fixed. Figure \ref{fig:spars} displays the results. Specifically, Figure \ref{subfig:spars_conv} displays the fluctuation amplitude CDF as a function of $\mu$ and $\plam$, such that $\plam$ varies, as $5 \leq m \leq 15$ varies, and $n=10^2$ and $\abs{\pnodes}/n = 0.2$ stay fixed. The color scale and red curve are defined as in Section \ref{subsec:size}. Figure \ref{subfig:spars_eval} displays the histogram of $\lambda_{\rm p}$ values for $m=5$, $10$, and $15$, with $10^4$ networks per $m$ value.
Figure \ref{subfig:spars_conv} confirms visually the analytical formula \eqref{stabcond2} for networks with $5 \leq m \leq 15$. As in Section \ref{subsec:size}, the simulations agree with \eqref{stabcond2}. Networks with $\plam$ above the red curve in Figure \ref{subfig:spars_conv} are predominantly yellow, hence agents asymptotically learn the false bias by $t=T$ for most coin toss sequences. Networks with $\plam$ below the red curve in Figure \ref{subfig:spars_conv} are predominantly blue, hence agents experience turbulent nonconvergence at $t=T$ for most coin toss sequences. \\

Figure \ref{subfig:spars_eval} shows how the distribution $p(\plam)$ depends on the attachment parameter $m$. Strictly speaking, sparsity and the attachment parameter $m$ are closely related but not exactly the same. A network is sparse, if the number of edges is much less than the maximum possible number of edges, i.e.\ $\abs{\mathcal{E}} \ll n(n-1)/2$ \cite{newman_networks_2018}. A network becomes denser as the number of edges approaches the maximum possible number of edges. In a BA network, one obtains $\abs{\mathcal{E}}=m + m(n-m)$\footnote{By the hand-shaking lemma \cite{newman_networks_2018}, the average degree of the network is equal to twice the attachment parameter, with $\sum_{i=1}^n d_i/n=2\abs{\mathcal{E}}/n\approx 2m$ for large $n$. Therefore, agents in networks with $5 \leq m \leq 15$ communicate with, on average, between $10$ and $30$ other agents.} \cite{newman_networks_2018}. Therefore, as $m$ increases with $n$ fixed, the sparsity of the network decreases, i.e. the network becomes denser. In Figure \ref{subfig:spars_eval}, $\langle \plam \rangle$ increases with $m$, with $\langle \plam \rangle=0.18$ for $m=5$ and $\langle \plam \rangle = 0.20 \approx \abs{\pnodes}/n$ for $m=15$. Additionally, the standard deviation decreases from $2.6\cross 10^{-2}$ for $m=5$ to $1.8 \cross 10^{-2}$ for $m=15$. The trend versus $m$ in $\langle \plam \rangle$ follows from $\min_i \sum_j W_{ij} \leq \rho(\vb{W}) \leq \max_i \sum_j W_{ij}$ \cite{meyer_matrix_2023}, which implies $\min_i d_{i,1}/d_i \leq \plam \leq \max_i d_{i,1}/d_i$
for $i \in \rnodes$. For a complete network, one has $d_{i,1}/d_i=\abs{\pnodes}/(n-1)$ for all $i \in \rnodes$ and hence $\plam \approx \abs{\pnodes}/n$ for large $n$. For $m=n-1$, the BA network is complete and approaches $\plam \rightarrow \abs{\pnodes}/n$. Hence the use of
\begin{equation}
    \label{meanf}
    \kldivbern{\theta_1}>-\log(1-\mu\abs{\pnodes}/n)
\end{equation}
as an approximation to \eqref{stabcond2} for large $m$ and $n$ distinguishes correctly when a typical BA network experiences turbulent nonconvergence. The right-hand side of \eqref{meanf} only depends on network parameters; it is independent of the statistical fluctuations caused by preferential attachment and partisan placement (i.e.\ it is deterministic).
\subsection{Partisan fraction}
\label{subsec:frac}
\begin{figure}[H]%
    \centering
    \begin{subfigure}[b]{0.49\linewidth}
        \centering
        \includegraphics[width=\linewidth]{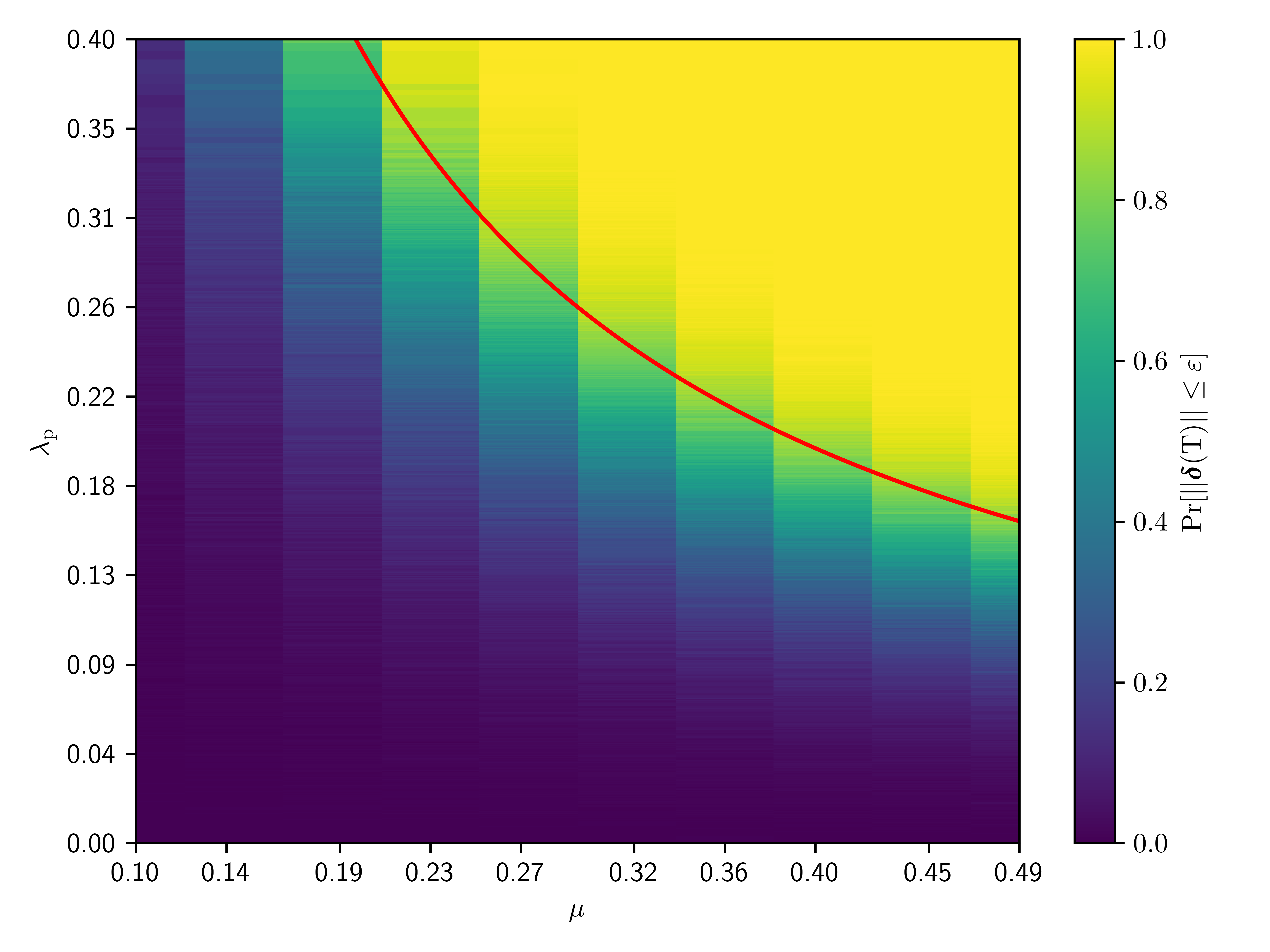}
        \caption{}
        \label{subfig:frac_conv}
    \end{subfigure}
    \begin{subfigure}[b]{0.49\linewidth}
        \centering
        \includegraphics[width=\linewidth]{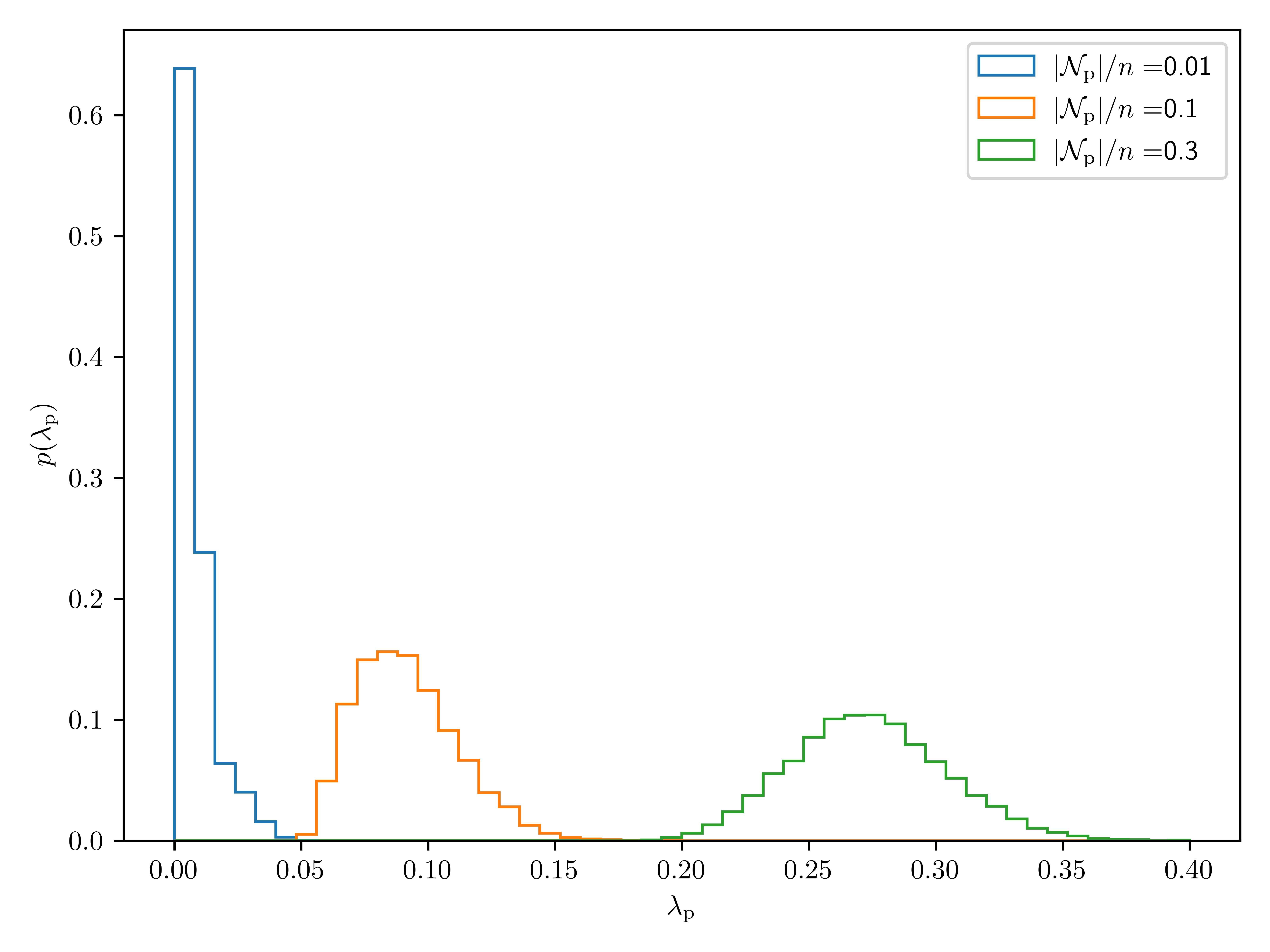}
        \caption{}
        \label{subfig:frac_eval}
    \end{subfigure}
    \caption{Effect of partisan fraction: as for Figure \ref{subfig:size_conv}, but with $0.01 \leq \abs{\pnodes}/n \leq 0.3$ and $n=10^2$ and $m=5$ fixed. (a) Heatmap of the fluctuation amplitude CDF. The red curve corresponds to the analytic instability threshold \eqref{stabcond2}. (b) Histogram of $\plam$ for $\abs{\pnodes}/n=0.01 \, \text{(blue curve)},0.1 \, \text{(orange curve)},0.3 \, \text{(green curve)}$.}
    \label{fig:frac}
\end{figure}

 In the last set of Monte Carlo simulations, we vary the partisan fraction $\abs{\pnodes}/n$ while holding $n$ and $m$ fixed. Figure \ref{fig:frac} displays the results. Specifically, Figure \ref{subfig:frac_conv} displays the fluctuation amplitude CDF as a function of $\mu$ and $\plam$, such that $\plam$ varies, as $0.01 \leq \abs{\pnodes}/n \leq 0.3$ varies, and $n=10^2$ and $m = 5$ stay fixed. The color scale and red curve are defined as in Section \ref{subsec:size}. Figure \ref{subfig:frac_eval} displays the histogram of $\plam$ values for $\abs{\pnodes}/n=0.01,0.1,$ and $0.3$, with $10^4$ networks per $\abs{\pnodes}/n$ value.
Figure \ref{subfig:frac_conv} confirms visually the analytic formula \eqref{stabcond2} for networks with $5 \leq m \leq 15$. As in Section \ref{subsec:size}, the simulations agree with \eqref{stabcond2}. Networks with $\plam$ above the red curve in Figure \ref{subfig:frac_conv} are predominantly yellow, hence agents asymptotically learn the false bias by $t=T$ for most coin toss sequences. Networks with $\plam$ below the red curve in Figure \ref{subfig:frac_conv} are predominantly blue, hence agents experience turbulent nonconvergence at $t=T$ for most coin toss sequences. \\

Figure \ref{subfig:frac_eval} shows how the distribution $p(\plam)$ depends on the partisan fraction $\abs{\pnodes}/n$. For $10^{-2} \leq \abs{\pnodes}/n \leq 10^{-1}$, we find $\langle \lambda_{\rm p} \rangle \approx \abs{\pnodes}/n$. In the above regime, $p(\plam)$ is unimodal but is not symmetric about the mode; it skews rightward to higher $\plam$. For $\abs{\pnodes}/n=0.3$, we find $\langle\plam\rangle = 0.27 < \abs{\pnodes}/n$, and $p(\plam)$ is unimodal and symmetric about the mode. The standard deviation increases as $\abs{\pnodes}/n$ increases, reaching $6.3\cross 10^{-3}$ for $\abs{\pnodes}/n=0.01$ and $2.9\cross 10^{-2}$ for $\abs{\pnodes}/n=0.3$. This, alongside the results in Sections \ref{subsec:size} and \ref{subsec:spars}, implies that the large-$n$ approximation \eqref{meanf} is accurate, when the network is large, dense, and has a low partisan fraction. Additionally, for $\mu=0.49$, the red curve in Figure \ref{subfig:frac_conv} predicts that the boundary between turbulent nonconvergence and asymptotic learning is for networks with $\plam \approx 0.15$. Given Figure \ref{subfig:frac_eval}, this means $\gtrsim 15\%$ of the network must be partisans for asymptotic learning to occur, even when partisans are at their most influential ($\mu=0.49$). \\

\section{Conclusion and social implications}
\label{sec:conclusion}
Partisans disrupt asymptotic learning about the political bias of a media organization in allies-only networks in two ways: some agents learn a false bias, or some agents never settle in their beliefs and experience turbulent nonconvergence \cite{bu_discerning_2023}. A single partisan is sufficient to disrupt a complete network, and a relatively small partisan fraction is sufficient to disrupt a scale-free BA network \cite{bu_discerning_2023,yildiz_binary_2013,mobilia_role_2007}. In this paper, we derive (and validate with Monte Carlo simulations) an analytic instability condition that distinguishes two modes of partisan disruption, in terms of the learning rate and key network properties, e.g.\ size, sparsity, and partisan fraction. The derivation relies on a two-state approximation, which reformulates the two-step update rule of an agent's belief PDF (based on independent observations and network-mediated peer pressure) into a system of nonlinear difference equations given by \eqref{eom}. From the stationary solutions of \eqref{eom}, we show that persuadable agents are disrupted from asymptotically learning the true bias if even one partisan exists, who believes in a false bias, and that the stationary solutions must also be a consensus of the persuadable subnetwork. When one or more partisans believe in a false bias, we derive an analytic instability condition, given by \eqref{stabcond2}, which demarcates asymptotic learning of the false bias from turbulent nonconvergence. Equation \eqref{stabcond2} expresses a balance between (i) the exogenous influence of the media organization's published outputs (idealized as coin tosses), captured by the KL divergence; and (ii) the endogenous influence of the partisans, captured by the learning rate $\mu$ and the network's pattern of connections, quantified through the smallest eigenvalue $\lambda_{\rm p}$ of the persuadable agent submatrix of the graph Laplacian $L_{ij}$, normalized by degree $d_i$. \\

We verify \eqref{stabcond2} by performing systematic Monte Carlo simulations and confirm that \eqref{stabcond2} agrees with the opinion dynamics observed empirically in previous work \cite{low_discerning_2022,bu_discerning_2023,low_vacillating_2022}. The instability condition is explored as a function of network size, sparsity, and partisan fraction. We find that the network is less likely to experience turbulent nonconvergence, as $\mu$ increases, $n$ decreases, $m$ increases, and $|{\cal N}_{\rm p}|/n$ increases. Table \ref{table2} summarizes the long-term behaviors observed for combinations of the network parameters $n$, $m$, and $\abs{\pnodes}/n$. We also find $\langle\plam\rangle \approx \abs{\pnodes}/n$ for large, dense BA networks with low partisan fraction. In this regime, the instability condition reduces approximately to \eqref{meanf}, a deterministic condition independent of statistical fluctuations associated with preferential attachment and partisan placement.
\begin{table}[H]%
    \centering
    \begin{tabular}{lcccc}
    \hline
                           & \begin{tabular}[c]{@{}c@{}} $\min m$\\ $\min |\mathcal{N}_{\rm{p}}|/n$\end{tabular} & \begin{tabular}[c]{@{}c@{}} $\min m$\\ $\max |\mathcal{N}_{\rm{p}}|/n$\end{tabular} & \begin{tabular}[c]{@{}c@{}} $\max m$\\ $\min |\mathcal{N}_{\rm{p}}|/n$\end{tabular} & \begin{tabular}[c]{@{}c@{}} $\max m$\\ $\max |\mathcal{N}_{\rm{p}}|/n$\end{tabular} \\ \hline
    $\min n$ & \begin{tabular}[c]{@{}c@{}}TN \\ (0.00)\end{tabular} & \begin{tabular}[c]{@{}c@{}}ALFB \\ (1.00)\end{tabular} & \begin{tabular}[c]{@{}c@{}}TN \\ (0.00) \end{tabular} & \begin{tabular}[c]{@{}c@{}}ALFB \\ (1.00) \end{tabular} \\
    $\max n$ & \begin{tabular}[c]{@{}c@{}}TN \\ (0.00)\end{tabular} & \begin{tabular}[c]{@{}c@{}}ALFB \\ (0.99) \end{tabular} & \begin{tabular}[c]{@{}c@{}}TN \\ (0.00) \end{tabular} & \begin{tabular}[c]{@{}c@{}}ALFB \\ (1.00) \end{tabular} \\ \hline
    \end{tabular}
    \caption{Long-term behavior predicted by the analytic instability condition \eqref{stabcond2} and confirmed by the numerical experiments in Section \ref{sec:montecarlo}. Possible outcomes are asymptotic learning of a false bias (ALFB) or turbulent nonconvergence (TN). ALFB corresponds to stability of the stationary solution $\hat{\pi}_{\rm{cons}}=1$, while TN corresponds to instability. There is no asymptotic learning of the true bias, as belief in the true bias is not a stationary solution. The numbers in parentheses are the fraction of networks with $\mathrm{Pr}[\norm{\vb*{\delta}(T)}\leq\varepsilon=0.05]=1$ out of the $10^2$ networks for each parameter combination. The eight parameter combinations correspond to the eight edge cases $(\min n = 30,\max n = 10^2) \otimes (\min m = 5, \max m = 15) \otimes (\min |{\cal N}_{\rm p}|/n = 0.02, \max |{\cal N}_{\rm p}|/n = 0.30)$, where $\otimes$ denotes a Cartesian (or alternatively outer) product. The table is constructed for $\mu= 0.25$.}
    \label{table2}
\end{table}
The disruption of asymptotic learning has been related elsewhere to structural balance theory (SBT) from the social sciences \cite{low_discerning_2022,bu_discerning_2023,antal_social_2006,davis_clustering_1967}. SBT categorizes networks as: (i) strongly balanced, if the networks can be partitioned into one or two clusters, within which all agents are cognitively cohesive or consonant ($A_{ij} > 0$); (ii) weakly balanced, if it can be partitioned into more than two clusters; and (iii) unbalanced, if it cannot be partitioned according to (i) or (ii). SBT predicts that balanced networks can asymptotically learn, and unbalanced ones cannot. In this paper, we study allies-only networks, which are always balanced. Therefore, the results of Section \ref{subsec:true} are consistent with SBT: persuadable agents in strongly balanced networks learn asymptotically. The weakly balanced, disconnected persuadable subnetworks in Section \ref{subsec:statpoint} are consistent with SBT, if each cluster contains only partisans who believe in the true bias. The two modes of partisan disruption distinguished by the instability condition \eqref{stabcond2} provide an interesting and partial counterpoint to SBT. Strongly and weakly balanced networks that do not satisfy \eqref{stabcond2} asymptotically learn a false bias, in agreement with SBT, whereas those that satisfy \eqref{stabcond2} undergo turbulent nonconvergence, thus contradicting SBT. However, partisans who believe in a false bias are cognitively dissonant with the exogenous influence of the coin, which is not encoded in $A_{ij}$ or SBT. This implies that SBT may be extended profitably to account for agents that are consonant endogenously but dissonant exogenously. \\

By applying the idealized model of a biased coin to the complicated real-world problem of media bias, we neglect a host of psychological, social, and political factors which are known to be salient in reality, e.g.\ cognitive biases like the hostile media effect, where media outputs are perceived as biased when they contradict one's preexisting attitudes \cite{eveland_impact_2003,perloff_three-decade_2015,lee_liberal_2005,southwell_roles_2007}. Nonetheless, with due reserve, we offer some brief thoughts about how the theoretical results in this paper apply to the formation of perceptions about media bias in human societies. One implication is that partisans define what can be learnt asymptotically by persuadable agents. This is reminiscent of the two-step flow model of media influence, where information passes from media organizations to certain members of a society, termed opinion leaders, who then disseminate their interpretation (accurately or otherwise) to the broader population \cite{southwell_roles_2007,tsang_opinion_2020}. In this framework, a society's perception of media bias is due to the endogenous influence of opinion leaders. In the hypodermic needle model of media influence, in contrast, media organizations dictate the perception of bias by communicating their outputs homogeneously across a society \cite{southwell_roles_2007,tsang_opinion_2020}. Interestingly, the update rule used in this paper aligns with the hypodermic needle model, as all agents observe identical coin tosses, yet partisans still emerge as opinion leaders in the spirit of the two-step flow model. Partisans can be understood as opinion leaders whose constancy is due to some purposeful intent or cognitive bias, such as the aforementioned hostile media effect which has been observed in empirical studies \cite{tsang_opinion_2020}. The results of this paper, for when partisans believe in a false bias, suggest that the mere presence of opinion leaders who misinform others destabilizes people's perception of bias by either convincing them of a falsehood or fomenting a climate of uncertainty by making them vacillate indefinitely. If the goal of malign opinion leaders is purely to disrupt discovery of the true bias, it can be achieved with low investment, i.e. with a low partisan fraction, and in networks of any size and sparsity. Trust of opinion leaders need not be fostered within a society, as disruption occurs irrespective of $\mu$. If the goal of malign opinion leaders is instead to teach a falsehood, the task is harder, with the results of Section \ref{sec:montecarlo} implying that $\gtrsim 15\%$ of the network are required to be partisans, even at their most influential ($\mu=0.49$). As the size and sparsity of the BA network increases, partisan influence decreases on average. This implies that to succeed in teaching a falsehood with minimal investment, partisans should target small, densely connected networks of persuadable agents. \\

The results in this paper open numerous avenues for future work. One important next step is to generalize the analytic calculations in this paper to opponents-only and mixed networks. Doing so is not easy. The presence of two additional nonlinearities in \eqref{eomgen} compared to \eqref{eom} makes solving for stationary solutions and evaluating the instability condition more difficult. Furthermore, the validity of the two-state approximation does not follow immediately from the simulations in Refs. \cite{low_discerning_2022,bu_discerning_2023,low_vacillating_2022}, unlike for an allies-only network. However, if the above difficulties can be overcome, there is the prospect of deriving new instability conditions of practical value. Monte Carlo simulations reveal that partisans disrupt opponents-only and mixed networks in interesting ways. In opponents-only networks, persuadable agents do not reach consensus due to antagonistic interactions but do achieve asymptotic learning individually, even in the presence of partisans \cite{bu_discerning_2023}. This indicates the presence of additional stationary solutions, which are stable and depend on an agent's position in the network. Like opponents-only networks, mixed networks do not reach consensus, but agents can exhibit turbulent nonconvergence and asymptotic learning of both the true bias and partisan beliefs \cite{bu_discerning_2023}. The additional stationary solutions may be unstable. \\

Another next step is to conduct a thorough analysis of the eigenvalue PDF $p(\lambda_{\rm p})$, with the aim of generating insights about how partisans should be optimally placed in the network to promote either asymptotic learning of a false bias, or turbulent nonconvergence. This optimization problem was flagged in Ref. \cite{bu_discerning_2023} and has been studied for a related but distinct eigenvalue in Refs. \cite{liu_optimizing_2021,zhou_optimization_2023,anaqreh_new_2024}. Additionally, we do not assume an explicit form of the likelihood in the paper, and $S(t)$ for all $t$ is an independent, identically distributed random variable. Hence the results in Sections \ref{sec:two-state} and \ref{sec:stability} apply to other opinion dynamics contexts beyond media bias, where agents attempt to infer an unknown parameter from a stochastic signal. Finally, going beyond the current model, one could make agents less homogeneous and let them learn at different rates and interpret $S(t)$ differently, i.e. $\mu \rightarrow \mu_i$ and $\lht{t}{\theta}\rightarrow P_i[S(t)|\theta]$ for $i\in\rnodes$. Such a generalization could encode intrinsic psychological differences, such as inherent partisanship, contrarian behavior, preexisting media attitudes, and other cognitive biases, as well as extrinsic environmental effects, like the role played by socioeconomic factors in disseminating media products. Many of these differences have been studied previously in connection with perceptions of media bias \cite{eveland_impact_2003,perloff_three-decade_2015,southwell_roles_2007,ho_role_2011}.

\section*{Acknowledgments}
We thank Nicholas Low for helping us to understand the model in Ref. \cite{low_discerning_2022}, Yutong Bu for key insights into partisan disruption from Ref. \cite{bu_discerning_2023}, and Yale Yauk for help with spectral graph theory. We also thank Yale Yauk for sharing a preliminary analysis of the opinion dynamics of an allies-only network updated according to step two of the update rule in Section \ref{subsec:step2}, i.e.\ without step one in Section \ref{subsec:step1}. We thank Justin Yu, Robin Evans, and Thippayawis Cheunchitra for additional discussions and editing. Jarra Horstman acknowledges the support of the ND Goldsworthy Scholarship. Andrew Melatos acknowledges funding from the Australian Research Council Centre of Excellence for Gravitational Wave Discovery (OzGrav) (CE230100016).

\bibliographystyle{elsarticle-num}
\bibliography{references}

\appendix
\section{Stability analysis}
\label{sec:appendix_a}
In this appendix, we derive formally the analytic instability condition \eqref{stabcond2}, which is the central result of the paper. \\

We analyze the linear stability of the nonlinear difference equation \eqref{eom} by considering perturbations about the stationary consensus, which is defined in Section \ref{subsec:statpoint}. That is, we write $\bel{i}{t}=\belcs+\delta_i(t)$, where $\delta_i(t)$ denotes the perturbation for agent $i \in \rnodes$ and is assumed to be small, viz.\ $\abs{\delta_i(t)} \ll 1$. Upon substituting into \eqref{eom}, we find that the perturbation evolves as
\begin{align}
    \delta_i(t+1) &= \bel{i}{t+1}-\belcs \\\label{perteom2}
    &= \sum_{j\in\rnodes} W_{ij} \bfunc{t}{\bel{j}{t}}+\frac{\mu d_{i,1}}{d_i} - \belcs \\\label{perteom3}
    &= \sum_{j\in\rnodes} W_{ij} \big\{\bfunc{t}{\delta_j(t)+\belcs}-\bfunc{t}{\belcs}\big\} \\\label{perteom4}
    &\approx f'[t,\belcs]\sum_{j \in \rnodes} W_{ij} \delta_j(t).
\end{align}
We pass from \eqref{perteom2} to \eqref{perteom3} by using $\belcs=\sum_{j \in \rnodes} W_{ij}\bfunc{t}{\belcs}+\mu d_{i,1}/d_i$ (see Section \ref{subsec:statpoint}), as $\belcs$ is a stationary solution, and from \eqref{perteom3} to \eqref{perteom4} by Taylor expanding. Upon writing vectors in bold type, the iterated version of \eqref{perteom4} after $t$ time steps is given by
\begin{equation}
    \label{perturb}
    \vb*{\delta}(t) = \vb{W}^t \vb*{\delta}(0)\prod_{n=0}^{t-1} f'(n,\belcs),
\end{equation}
where $\vb{W}$ is the submatrix of $\vb{I}-\mu \vb{D}^{-1}\vb{L}$ for persuadable agents, and we write $\vb{D}_{ij}=\mathrm{diag}{\left(d_1,\ldots,d_{\abs{\rnodes}}\right)}$ and $\vb*{\delta}(t)=[\bel{1}{t}-\belcs,\ldots,\bel{\abs{\rnodes}}{t}-\belcs]^\top$, where the superscript $\top$ symbolizes the matrix transpose. \\

The stationary consensus is asymptotically stable, if we have $\lim_{t\rightarrow\infty}\norm{\vb*{\delta}(t)}=0$, and is unstable otherwise. That is, the perturbations for every agent are deemed to decay and yield stability, if the length of the perturbation vector, or fluctuation amplitude, as measured by an arbitrary vector norm $\norm{\cdot}$, approaches zero. It is shown in Theorem 4.33 of \cite{elaydi_introduction_2005} that $\lim_{t\rightarrow\infty}\norm{\vb*{\delta}(t)} = 0$ and \eqref{perturb} imply that the stationary consensus is stable given the exact difference equation \eqref{eom} and is unstable otherwise. Theorem 4.33 is independent of the choice of norm. \\

The formal notion of asymptotic stability differs slightly from that discussed in the previous paragraph, as the time dependence in \eqref{perturb} is due to the sequence of Bernoulli random variables. This amounts to the limit $t\rightarrow\infty$ being taken probabilistically, so that time-averages converge to a mean value taken with respect to the Bernoulli random variable through the Law of Large Numbers. Formally speaking, one says that the instability condition holds with probability one, which means that when the ensemble of all possible coin toss sequences is considered, the probability of selecting a pathological finite sequence that contradicts the instability condition goes to zero as $t\rightarrow\infty$. To show asymptotic stability, we bound $\norm{\vb*{\delta}(t)}$ from above by
\begin{align}
    \label{pertupper1}
    \norm{\vb*{\delta}(t)} &= \norm{\vb{W}^t\vb*{\delta}(0)}\prod_{n=0}^{t-1}\abs{f'(n,\belcs)} \\\label{pertupper2}
    &\leq \norm{\vb{W}^t}\norm{\vb*{\delta}(0)}\prod_{n=0}^{t-1}\abs{f'(n,\belcs)} \\\label{pertupper3}
    &= \norm{\vb*{\delta}(0)}\exp{t\left[\log \norm{\vb{W}^t}^{\frac{1}{t}}+\frac{1}{t}\sum_{n=0}^{t-1} \log\abs{f'(n,\belcs)}\right]}.
\end{align}
We pass from \eqref{pertupper1} to \eqref{pertupper2} using $\norm{\vb{A}\vb{x}}\leq \norm{\vb{A}}\norm{\vb{x}}$ for any matrix-vector product, where $\norm{\vb{A}}$ is the matrix norm induced by the vector norm \cite{meyer_matrix_2023}. We pass from \eqref{pertupper2} to \eqref{pertupper3} by writing the products in \eqref{pertupper2} as the exponential of a sum of logarithms. The fluctuation amplitude grows or decays, if the exponent in \eqref{pertupper3} is positive or negative, respectively, as $t\rightarrow\infty$. \\

To take the limit $t\rightarrow\infty$, we note two formulas. First, Gelfand's formula implies $\lim_{t\rightarrow\infty} \norm{\vb{A}^t}^{1/t}=\rho(\vb{A})$ for any matrix $\vb{A}$, where $\rho(\vb{A})=\max_i \abs{\lambda_i(\vb{A})}$ is the spectral radius \cite{meyer_matrix_2023}. Second, as $f'(t,\belcs)$ is a function of a Bernoulli random variable, $\{\log \abs{f'(t,\belcs)}\}_{t=0}^\infty$ is a sequence of independent, identically distributed random variables. Hence, their average converges to $\mathbb{E}\big[\log \abs{f'(t,\belcs)}\big]$ by the Law of Large Numbers, where the expectation $\mathbb{E}$ is taken within respect to the Bernoulli distribution \cite{lehmann_testing_2005}. Applying the two formulas above to \eqref{pertupper3}, we obtain
\begin{equation}
    \label{cond}
    \lim_{t\rightarrow\infty}\norm{\vb*{\delta}(t)} = \begin{cases}
        0 & \text{if} \,\, \mathbb{E}\big[\log \abs{f'(t,\belcs)}\big]<-\log \rho(\vb{W}) \\
        \infty & \text{if} \,\, \mathbb{E}\big[\log \abs{f'(t,\belcs)}\big]>-\log \rho(\vb{W})
    \end{cases}
\end{equation}
Evaluating \eqref{cond} for the stationary consensus $\belcs=1$ gives
\begin{align}
    \mathbb{E}\big[\log \abs{f'(t,1)}\big] &= -\mathbb{E}\big[\log \bfac{t}\big] \\
    &= \mathbb{E}\big\{\log \lht{t}{\theta_2}\big\} -  \mathbb{E}\big\{\log \lht{t}{\theta_1}\big\} \\
    \label{deriv}
    &= \kldivbern{\theta_1}-\kldivbern{\theta_2},
\end{align}
where $\mathrm{KL}$ is the Kullback-Leibler divergence
\begin{equation}
    \label{kldivappen}
    \kldivbern{\theta}=\sum_{S(t)\in\{0,1\}} \bern \log{\left\{\frac{\bern}{\lht{t}{\theta}}\right\}}.
\end{equation}
The sum in \eqref{kldivappen} is over heads [$S(t)=1$] and tails [$S(t)=0$] at time $t$. Likewise, evaluating \eqref{cond} for $\belcs=0$ gives
\begin{align}
    \mathbb{E}\big[\log \abs{f'(t,0)}\big] &= \mathbb{E}\big[\log \bfac{t}\big] \\
    &= \mathbb{E}\big\{\log \lht{t}{\theta_1}\big\} -  \mathbb{E}\big\{\log \lht{t}{\theta_2}\big\} \\
    \label{deriv2}
    &= \kldivbern{\theta_2}-\kldivbern{\theta_1},
\end{align}
Equations \eqref{deriv} and \eqref{cond} combine with the assumption $\theta_2=\bias$ to give the instability condition \eqref{stabcond2} in Section \ref{sec:stability}, which is the central result of the paper. Equation \eqref{deriv} is a well-known result in the Bayesian asymptotic literature, however, it is proven with different mathematical tools \cite{shalizi_dynamics_2009}. It holds in other situations too, such as for more complicated likelihood functions and when $S(t)$ is continuous-valued and correlated through time \cite{shalizi_dynamics_2009,walker_bayesian_2013,chatterjee_short_2020}. \\

The stability results above do not depend on the chosen norm. The norm used to analyze the simulations in Section \ref{sec:montecarlo} follows from considering the total variation distance,
\begin{align}
    \label{tv2}
    \norm{\opin{i}{\theta}{t}-\hat{x}_{\mathrm{cons}}(\theta)}_{\mathrm{TV}} &= \max_{\theta\in \topic} \abs{\opin{i}{\theta}{t}-\hat{x}_{\mathrm{cons}}(\theta)} \\\label{tv}
    &= \frac{1}{2}\sum_{\theta \in \topic} \abs{\opin{i}{\theta}{t}-\hat{x}_{\mathrm{cons}}(\theta)}.
\end{align}
We write $\topic = \{\theta_1,\ldots,\theta_k\}$ as defined in Section \ref{sec:model} and pass from \eqref{tv2} to \eqref{tv} using a standard result that holds for any two PDFs \cite{cover_elements_2005}. Equation \eqref{tv} can be interpreted as measuring the largest difference in belief, or disagreement, with the stationary consensus that agent $i$ maintains at time $t$ across all $\theta$. Tracking the agent with the greatest disagreement gives the $L_\infty$-norm for the two-state approximation,
\begin{align}
    \max_{i\in\rnodes} \norm{\opin{i}{\theta}{t}-\hat{x}_{\mathrm{cons}}(\theta)}_{\mathrm{TV}} &= \max_{i\in\rnodes} \abs{\bel{i}{t}-\belcs} \\\label{tvtwo}
    &= \norm{\vb*{\delta}(t)}.
\end{align}
Equation \eqref{tvtwo} is used throughout Section \ref{sec:montecarlo}.
\end{document}